\documentclass[twocolumn]{aastex631}

\received{January 1, 2022}
\accepted{for publication in ApJ – March 2, 2022}
\usepackage{textgreek}

\shorttitle{Identification and spectroscopic characterization of 128 new Herbig stars}
\shortauthors{Vioque et al.}
\begin{document}

%\title{Template \aastex Article with Examples: v6.3\footnote{Released on June, 10th, 2019}}

\title{Identification and spectroscopic characterization of 128 new Herbig stars\footnote{Tables B1, B2, and B3 are available at the CDS in their entirety in machine-readable form via \href{http://vizier.u-strasbg.fr/viz-bin/VizieR?-source=J/ApJ/930/39}{http://vizier.u-strasbg.fr/viz-bin/VizieR?-source=J/ApJ/930/39} or via \href{https://cdsarc.unistra.fr/ftp/J/ApJ/930/39}{https://cdsarc.unistra.fr/ftp/J/ApJ/930/39}.}\footnote{Based on observations collected at the European Southern Observatory (ESO) under ESO programme 0104.C-0937(A) and 0104.C-0937(B), observations made in the Observatorios de Canarias del IAC, programme 103-INT13/19B, and observations collected at the Centro Astronómico Hispano-Alemán (CAHA) at Calar Alto from 2019-04-Sept to 2019-06-Sept.}}

\correspondingauthor{Miguel Vioque}
\email{miguel.vioque@alma.cl}

\author[0000-0002-4147-3846]{Miguel Vioque}
\affiliation{Joint ALMA Observatory, Alonso de Córdova 3107, Vitacura, Santiago 763-0355, Chile}%\email{miguel.vioque@alma.cl}
\affiliation{National Radio Astronomy Observatory, 520 Edgemont Road, Charlottesville, VA 22903, USA}
\affiliation{School of Physics and Astronomy, Sir William Henry Bragg Building, University of Leeds, Leeds LS2 9JT, UK}

\author[0000-0001-7703-3992]{René D. Oudmaijer}
\affiliation{School of Physics and Astronomy, Sir William Henry Bragg Building, University of Leeds, Leeds LS2 9JT, UK}

\author{Chumpon Wichittanakom}
\affiliation{School of Physics and Astronomy, Sir William Henry Bragg Building, University of Leeds, Leeds LS2 9JT, UK}
\affiliation{Department of Physics, Faculty of Science and Technology, Thammasat University, Rangsit Campus, Pathum Thani 12120, Thailand}

\author[0000-0002-0233-5328]{Ignacio Mendigutía}
\affiliation{Centro de Astrobiología (CSIC-INTA), Departamento de Astrofísica, ESA-ESAC Campus, 28691 Madrid, Spain}

\author[0000-0002-6923-3756]{Deborah Baines}
\affiliation{Quasar Science Resources for ESA-ESAC, ESAC Science Data Centre, PO Box 78, 28691 Villanueva de la Cañada, Madrid, Spain}

\author[0000-0002-6648-2968]{Olja Panić}
\affiliation{School of Physics and Astronomy, Sir William Henry Bragg Building, University of Leeds, Leeds LS2 9JT, UK}

\author[0000-0002-0756-9836]{Daniela Iglesias}
\affiliation{School of Physics and Astronomy, Sir William Henry Bragg Building, University of Leeds, Leeds LS2 9JT, UK}

\author[0000-0002-1575-680X]{James Miley}
\affiliation{Joint ALMA Observatory, Alonso de Córdova 3107, Vitacura, Santiago 763-0355, Chile}
\affiliation{National Astronomical Observatory of Japan, Alonso de Córdova 3788, 61B Vitacura, Santiago, Chile}

\author[0000-0002-9127-5522]{Ricardo Pérez-Martínez}
\affiliation{Ingeniería de Sistemas para la Defensa de España (Isdefe) for  ESA-ESAC. PO Box 78, 28691 Villanueva de la Cañada, Madrid, Spain}

\begin{abstract}

%\textbf{Abstract under construction - 250 word limit}

%low-resolution

We present optical spectroscopy observations of 145 high-mass pre-main sequence candidates from the catalogue of \citet{2020A&A...638A..21V}. From these, we provide evidence for the Herbig nature of 128 sources. This increases the number of known objects of the class by $\sim50\%$. We determine the stellar parameters of these sources using the spectra and Gaia EDR3 data. The new sources are well distributed in mass and age, with 23 sources between $4$-$8$ M$_{\odot}$ and 32 sources above $8$ M$_{\odot}$. Accretion rates are inferred from H\textalpha{} and H\textbeta{} luminosities for 104 of the new Herbigs. These accretion rates, combined with previous similar estimates, allow us to analyze the accretion properties of Herbig stars using the largest sample ever considered. We provide further support to the existence of a break in accretion properties at $\sim3$-$4$ M$_{\odot}$, which was already reported for the previously known Herbig stars. We re-estimate the potential break in accretion properties to be at $3.87^{+0.38}_{-0.96}$ M$_{\odot}$. As observed for the previously known Herbig stars, the sample of new Herbig stars independently suggests intense inner-disk photoevaporation for sources with masses above $\sim7$ M$_{\odot}$. These observations provide robust observational support to the accuracy of the \citet{2020A&A...638A..21V} catalogue of Herbig candidates.

\end{abstract}

%% Keywords should appear after the \end{abstract} command. 
%% See the online documentation for the full list of available subject
%% keywords and the rules for their use.
\vspace{-50 pt}

%% From the front matter, we move on to the body of the paper.
%% Sections are demarcated by \section and \subsection, respectively.
%% Observe the use of the LaTeX \label
%% command after the \subsection to give a symbolic KEY to the
%% subsection for cross-referencing in a \ref command.
%% You can use LaTeX's \ref and \label commands to keep track of
%% cross-references to sections, equations, tables, and figures.
%% That way, if you change the order of any elements, LaTeX will
%% automatically renumber them.
%%
%% We recommend that authors also use the natbib \citep
%% and \citet commands to identify citations.  The citations are
%% tied to the reference list via symbolic KEYs. The KEY corresponds
%% to the KEY in the \bibitem in the reference list below. 

\section{Introduction} \label{sec:intro}

The group of intermediate- to high-mass pre-main sequence (PMS) sources (M\textgreater{}$1.5\,\textrm{M}_{\odot}$) play a particularly important part in understanding the differences in the formation of, and accretion of matter onto, low- and high-mass stars. Whereas the formation of low-mass stars is widely accepted to be due to magnetically controlled accretion (\citealp{2007prpl.conf..479B}), higher mass stars are generally non-magnetic, and it would be expected that this scenario does not apply to them (see \citealp{2020Galax...8...39M}). Recent studies (e.g. \citealp{2020MNRAS.493..234W,2022ApJ...926..229G}) have provided evidence for the change in accretion mechanism happening at around $4$ M$_{\odot}$.

%Mm-wavelength surveys of hundreds of low-mass PMS stars in nearby star forming regions have provided important advances in our understanding of protoplanetary disks and planet formation (e.g. \citealp{2016ApJ...828...46A}).

In addition, the protoplanetary disks of intermediate- to high-mass PMS stars show significant differences with respect to the disks around low-mass stars. The more luminous stars promptly photoevaporate their inner disks with FUV photons (\citealp{2021ApJ...909..109K}), causing large inner cavities. The impact this has on the thermal and chemical evolution of the disks may significantly impact the formation and evolution of planets (\citealp{2017MNRAS.467.1175P}; \citealp{2021MNRAS.500.4658M}). Mm-wavelength observations have also shown that the disks around intermediate-mass stars are more massive than those around low-mass stars (\citealp{2013ApJ...771..129A, 2016ApJ...831..125P, 2022A&A...658A.112S}) and have a higher fraction of detected structures in the dust disk (\citealp{2021AJ....162...28V,2022A&A...658A.112S}). Furthermore, there is evidence pointing toward particular disk structures being favored in more massive objects (e.g., spirals have been mostly found in early spectral type stars, \citealp{2018A&A...620A..94G}). Hence, it has been theorized that they have a higher incidence of giant planets (e.g. \citealp{2015A&A...574A.116R, 2021AJ....161...33V}). Therefore, the disks around intermediate- to high-mass PMS stars play a key role for understanding the structures and evolution processes of protoplanetary disks, and their link to planet formation.

The intermediate- to high-mass PMS regime comprises different types of young stellar objects (YSOs). The Herbig Ae/Be group contains stars at the latest stages of pre-main sequence evolution. In total, around 255 Herbig Ae/Be sources have been historically considered and studied (e.g. \citealp{2018A&A...620A.128V}), although a smaller fraction has been properly characterized and is free of contaminants (see the seminal papers of \citealp{1994A&AS..104..315T}, \citealp{2003AJ....126.2971V}, and \citealp{2004AJ....127.1682H}). Recent compilations and studies about the general properties of Herbig Ae/Be stars can be found in \citet{2018A&A...620A.128V} and \citet{2021A&A...650A.182G}. The cooler predecessors of the Herbig Ae/Be stars are occasionally referred to as Intermediate-Mass T Tauris (IMTTs). The defining difference between both groups is arbitrary in the literature and different thresholds in spectral type and mass have been used (e.g. \citealp{2004AJ....128.1294C, 2016ApJ...825..125P, 2019A&A...622A..72V, 2021AJ....162..153N, 2021A&A...652A.133V}). \citet{2021A&A...652A.133V} compiled most IMTTs within 500 pc, describing the general properties of a sample of 49 sources. Higher-mass PMS objects (M\textgreater{}$8$-$10$ M$_{\odot}$) are generally called Massive Young Stellar Objects (MYSOs). However, the difference between the former two categories and the MYSOs is also ambiguous, often depending on the optical visibility of the sources. The most comprehensive catalogue of MYSOs to date can be found in \citet{2013ApJS..208...11L} with several hundred sources, although only a very small fraction of those have been characterized in great detail (e.g. \citealp{2021A&A...648A..62F}; \citealp{2021A&A...654A.109K}).

%\textbf{Hence, it is unclear how many IMTTs have been properly characterised}.
%\textbf{\citealp{2021AJ....162..153N}, using infrared and X-ray observations described 77 IMTTs in the Carina Nebula}. 

A large caveat of all the aforementioned results is that there is no homogeneous or complete survey of intermediate- to high-mass PMS stars. The existing samples are small heterogeneous collections of often randomly discovered objects, with a non-negligible amount of contaminants. This is a direct consequence of the Initial Mass Function (IMF): there are few nearby high-mass stars in any given molecular cloud (e.g., there are three stars with $M>1.5$ M$_{\odot}$ in the ALMA survey of Lupus, \citealp{2016ApJ...828...46A}). Any global conclusions that have been drawn are therefore subject to an unknown bias. Furthermore, as more massive stars evolve faster, the younger, high-mass PMS stars have been barely considered by the vast majority of studies (e.g. in the samples of \citealp{2021AJ....161...33V} and \citealp{2022A&A...658A.112S} all sources are older than $\sim4$ Myr). This has important consequences on the conclusions obtained so far for these objects. For example, at those late stages, planet formation is mostly over in the low-mass regime (\citealp{2021MNRAS.501.2934C}), so it is probable that in general we have not been tracing actively planet-forming stars.

%In addition, the vast majority of studies dedicated to massive PMS stars focus on the lower mass end of the class.

%Hence, most of what is known about intermediate- to high-mass star formation is based on individual observations of several evolved stars in a narrow mass range ($\sim1.5$-$2.5$ M$_{\odot}$).

%although the observations of protoplanetary disks around high-mass stars are scarce, 

To properly investigate the evolution of intermediate- and high-mass PMS stars, what is needed is a well-selected sample covering a large range in age and mass. \citet{2020A&A...638A..21V} produced a large homogeneous catalogue of a few thousand intermediate- to high-mass PMS (`Herbig star'\footnote{In this work we call all YSOs with masses above $1.5$ M$_{\odot}$ `Herbig stars'. We elected this nomenclature because most of the sources identified and described in this work clearly belong the Herbig Ae/Be group, although some others might better fit within the IMTTs or MYSOs regimes.}) candidates by combining large scale photometric surveys (Gaia, 2MASS, WISE, IPHAS, and VPHAS+; covering from the optical to the mid-IR and including H$\alpha$ photometry) with machine learning techniques. 

In this paper, we present spectroscopic observations for 145 Herbig candidates from the catalogue of \citet{2020A&A...638A..21V}. We discuss the results of the observations and present a comprehensive list of 128 newly confirmed Herbig detections. We start the paper with a description of the observations in Sect. \ref{s_observations}. For each observed object, we characterize the extinction and the stellar parameters in Sect. \ref{S_stellar_parameters}. After discarding some contaminants in Sect. \ref{S_contaminants}, in Sect. \ref{S_accre} we present accretion rates for the fraction of the sources with H\textalpha{} or H\textbeta{} emission line measurements. In Sect. \ref{S_PMS_nature} the Herbig nature of the observed sources is assessed. In Sect. \ref{S_accretion_analysis} the derived accretion rates are analyzed, and we compare them with the accretion rates of previously known Herbig stars. We discuss our results in Sect. \ref{S_discussion} and conclude in Sect. \ref{S_Conclusions}.

%llllcccc
\begin{deluxetable*}{lllllcccc}
%\tablewidth{10pt}
\tablecolumns{8}
\tablecaption{Observing dates and instrumental setups for the three observing runs; including telescope, spectrograph, CCD detector, grating or grism used, spectral range in \AA{},  reciprocal dispersion in $\text{\AA{}}/\text{pixel}$, and spectral resolution in \AA{}.\label{Table_appexA_1}}
\tablehead{
Date ranges & Telescope&	Instrument&	Detector&	Grating/Grism &	Range &	Dispersion&	Resolution & Number\\
& &					&					&				&				(\AA)&		(\AA\,pixel$^{-1}$)&	(\AA)& of sources	}
\startdata
4th-7th Aug 2019 & INT&		IDS&			EEV10&		R900V&	3600–5000&	0.63&	$\sim1.3$& 56\\
8th-11th Aug 2019 & INT&			IDS&			RED+2&		R1200R&	5700–6700&	0.52&	$\sim1.0$&	 56\\
3rd-5th Sep 2019 & CAHA2.2m &	CAFOS&	 SITe-1d &	B-100&		3200–5800&	2.0&		$\sim4.0$& 39\\
11th-13th Mar 2020 & NTT&	EFOSC2&	CCD\#40&		G7&	3300–5300&	0.96&	$\sim7.4$&	 50\\
17th-20th Mar 2020 & NTT&			EFOSC2&	CCD\#40&		G20&	6000–7200&	0.55&	$\sim3.5$& 48\\
\enddata
%\end{tabular}
\tablecomments{The signal-to-noise ratio of these spectra are typically in the order of 100. A total of 145 different sources were observed.}
\end{deluxetable*}

%We characterise spectral types and hence obtain total extinction values for a subset of observable massive PMS objects, allowing them to be placed in the HR diagram, which is essential to derive accurate stellar parameters.

%This way young objects can be compared with more evolved objects in a homogeneous sample. Such a sample needs to cover a wide range of ages, including the earliest stages of PMS evolution. Therefore, observations of the more massive and younger high-mass PMS stars are necessary to complement previous results and capture the full picture of the protoplanetary disks around these objects. 

%We will accurately determine the line-of-sight extinction and stellar parameters such as temperature, luminosity, mass, age, radius, rotation speed and metallicity. We also determine the accretion rates of this set of objects, and study their inner disks.

\section{Observations} \label{s_observations}

A total of 145 Herbig candidates from the \citet{2020A&A...638A..21V} catalogue were observed in low- to medium-resolution optical spectroscopy during three different observing runs. Hence, this is a pilot study of the large catalogue of 2226 intermediate-  to  high-mass  PMS candidates of \citet[only counting those sources with $\sigma(\varpi)/\varpi\leq0.2$]{2020A&A...638A..21V}. These 145 sources were selected because their absolute magnitudes suggest that their stellar masses cover the Herbig mass regime in a representative manner. In order to obtain precise stellar parameters, we targeted sources with accurate parallaxes ($\varpi$, all but five have $\sigma(\varpi)/\varpi\leq0.1$). This is because the parallax error dominates the uncertainty of the stellar parameters when these are obtained from the locations of the objects in the HR diagram (see e.g. \citealp{2018A&A...620A.128V}).

%These were selected from the \citet{2020A&A...638A..21V} catalogue because of their expected high-mass as traced by their absolute magnitude corrected from interstellar extinction (presented in \citealp{2020A&A...638A..21V}). As the interstellar extinction is a lower limit, the lower limits to their masses should ensure they are massive objects indeed. 

%the magnitude range traced in our observations (15.5\textgreater{}G\textgreater{}$12$ mag

The signal-to-noise ratio of these spectra are typically on the order of 100. Observing dates and instrumental setups are detailed in Table~\ref{Table_appexA_1}. The three observing runs can be summarised as follows:

%The number of sources observed in each observing run are tabulated in Table \ref{Table_observations}, together with the reciprocal dispersion and the spectral resolution of each configuration used:

\begin{itemize}
    \item 56 Herbig candidates were observed with the Intermediate Dispersion Spectrograph (IDS) instrument which is at the Cassegrain focus of the $2.54$-metre Isaac Newton Telescope (INT). The INT is located at the Roque de los Muchachos Observatory in the island of La Palma, Spain. Two different configurations were used. One block used the R900V diffraction grating, which covers the $\sim3600-5000$ \r{A} spectral range.  The other block used the R1200R grating ($\sim5700-6700$ \r{A}).
    
   % In two blocks of four nights each
    %the EEV10 CCD detector together with
%with the Red+2 CCD detector

%This setting has a reciprocal dispersion of $0.63$ \r{A}/pixel and a spectral resolution of $\sim 1.3$ \r{A} (or resolving power $R\sim3400$).
%This setting has a reciprocal dispersion of 0.52 \r{A}/pixel and a spectral resolution of $\sim 1.0$ \r{A} (or resolving power $R\sim6000$).

    \item 39 Herbig candidates were observed with the Calar Alto Faint Object Spectrograph (CAFOS) at the RC focus of the Calar Alto 2.2-metre telescope (CAHA2.2m) in Calar Alto Observatory, Spain. We employed the B-100 grism ($\sim3200-5800$ \r{A}).
    
    %and SITe-1d 2Kx2K CCD chip
    
     %which produces a reciprocal dispersion of $2.0$ \r{A}/pixel and a spectral resolution of $\sim 4.0$ \r{A} (or resolving power $R\sim1100$)

    \item 50 Herbig candidates were observed with the ESO Faint Object Spectrograph and Camera (v.2) or EFOSC2 in two settings. EFOSC2 is installed at the Nasmyth B focus of the 3.58-metre New Technology Telescope (NTT) at La Silla Observatory, Chile. The first block used the G7 diffraction grism ($\sim 3300-5300$ \r{A}) and the second block used the G20 grism ($\sim 6000-7200$ \r{A}).
    
    %with CCD\#40
    
    %The G7 grism ($\sim 3300-5300$ \r{A}) provides a reciprocal dispersion of 0.96 \r{A}/pixel and a spectral resolution of $\sim 7.4$ \r{A} (or resolving power $R\sim600$). The higher resolution G20 grism ($\sim 6000-7200$ \r{A}) provides a reciprocal dispersion of 0.55 \r{A}/pixel and a spectral resolution of $\sim 3.5$ \r{A} (or resolving power $R\sim1900$).
\end{itemize}

A log of the observations is presented in Table \ref{Table_log_observations}. Bias, flat and arc frames were taken each night for the reduction of the observations. Standard procedures were used in order to process the data, which was reduced using the Image Reduction and Analysis Facility (\texttt{IRAF}). We started with bias subtraction. Next, flat field division was used to correct the pixel-to-pixel variation of the CCD signal. Then, one-dimensional spectra were extracted and sky subtracted from science frames. Last, arc frames of Cu-Ar and Cu-Ne comparison lamps were used to obtain the wavelength calibrated normalized spectra. 

%More information about the data reduction process can be found in Appendix \ref{App_obs}. 

\subsection{Comments on the blue spectral range}\label{S_blue_range}

\begin{figure}
\centering
\includegraphics[width=\columnwidth]{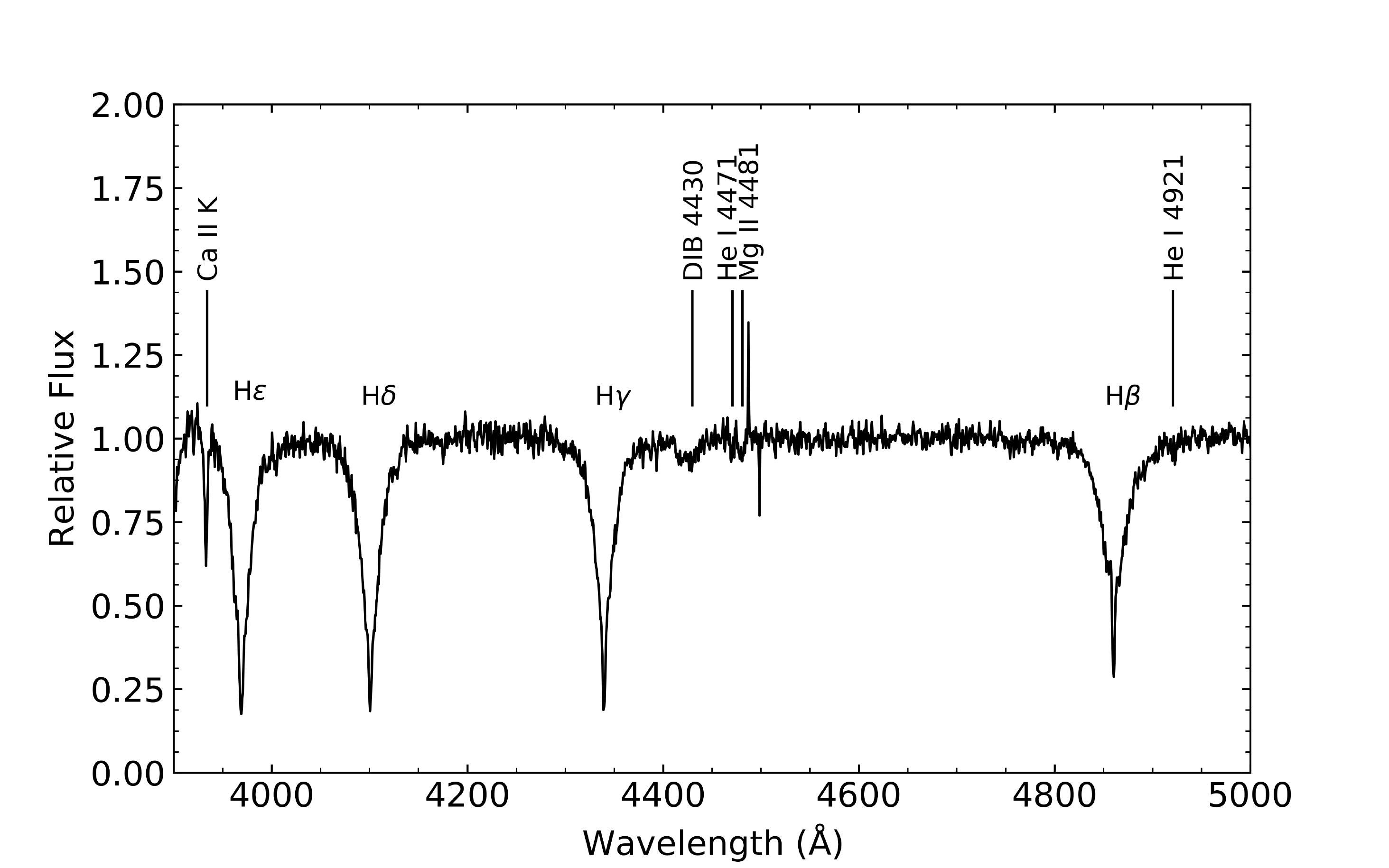}
\includegraphics[width=\columnwidth]{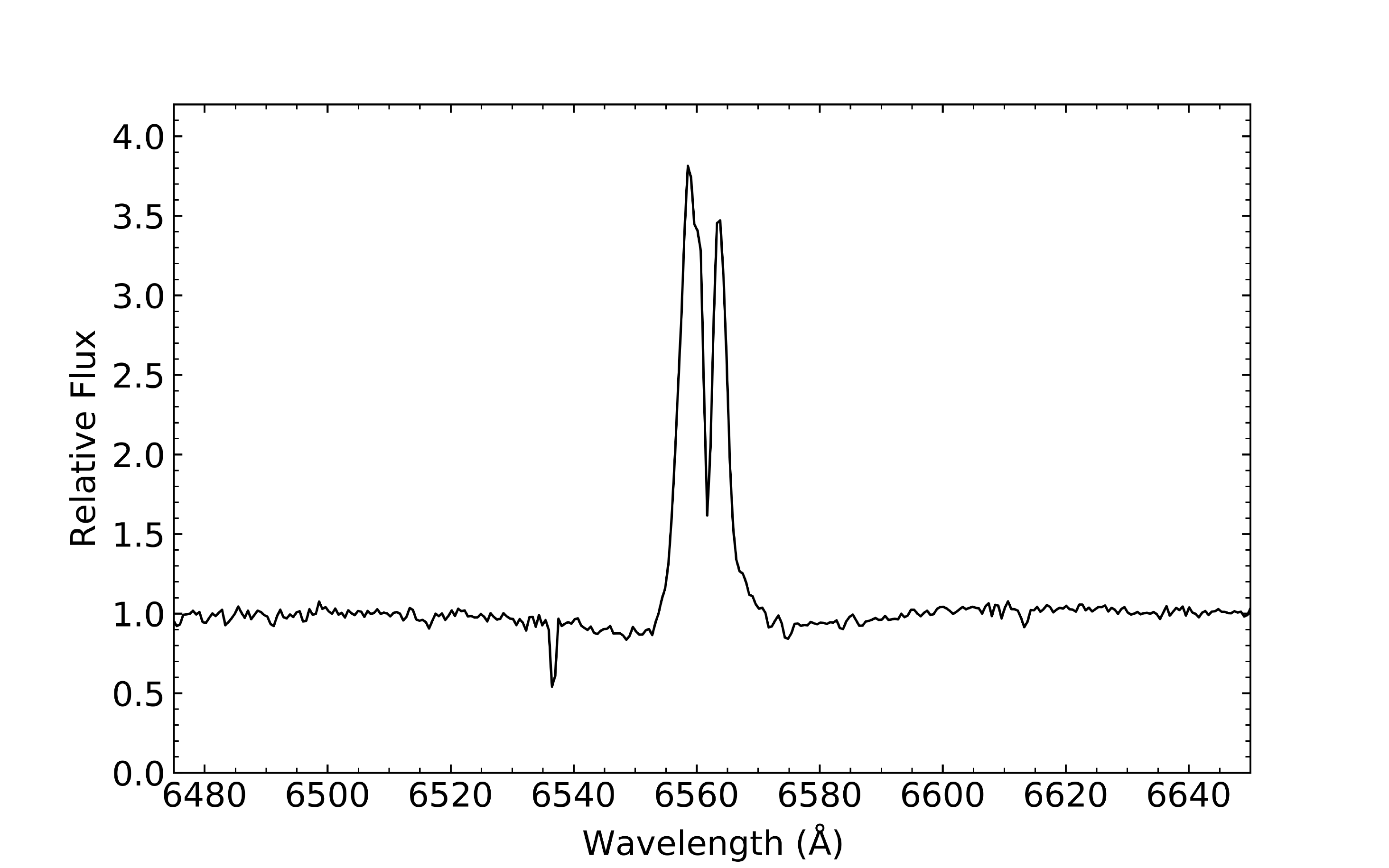}
\caption{Example normalized spectra of VOS 140 (spectral type B9.5), observed at INT. \textit{Top:} Blue spectral range covered by our observations. \textit{Bottom:} Portion of the red spectral range covered by our observations centered around the H\textalpha{} line.}
\label{fig:Example spectra}
\end{figure}

The blue spectral region considered in the three observing runs ($\sim3300-5400$ \r{A}) covers the main wavelength range to determine spectral types. An example of the normalized spectra obtained in this region is shown in Fig. \ref{fig:Example spectra}. This region is especially useful for the earlier spectral type stars such as A and B stars, as their spectrum in the wavelength range beyond $5000$ \r{A} is fairly line free. This region also covers the H\textbeta{} line, which is a common tracer of circumstellar activity in YSOs.

%It also covers the Balmer jump 

The chosen grisms allow for efficiently obtaining spectral types and effective temperatures (T\textsubscript{eff}). These are tabulated in Table \ref{Table_stellar_param}. For some objects the temperature was estimated directly from model fitting the spectra (see \citealp{2020MNRAS.493..234W}). Otherwise spectral types were obtained by comparison with model spectra (BOSZ models, \citealp{2017AJ....153..234B}) and published spectral standards (Digital Atlas by Gray\footnote{\dataset[https://ned.ipac.caltech.edu/level5/Gray/frames.html]{https://ned.ipac.caltech.edu/level5/Gray/frames.html}}), and the T\textsubscript{eff} in Table \ref{Table_stellar_param} are the values that correspond to those spectral types according to \citet{2013ApJS..208....9P}. In this latter case, the T\textsubscript{eff} uncertainties are of one sub-spectral type. The procedure followed for each object is detailed in Table \ref{Table_stellar_param}. T\textsubscript{eff} could not be estimated for VOS 2164 and VOS 603 because of strong emission line spectra (see Sect. \ref{S_contaminants}). `VOS' names refer to sources from the catalogue of \textit{`Vioque, Oudmaijer, Schreiner, et al. 2020'}.

%and thus no spectral type was derived for those sources

%Examples of the normalised spectra obtained in this spectral range are shown in Fig. \ref{fig_observed_spectra} in the top panels.

%For those observed sources for which spectral types were determined we took one spectral sub-type as uncertainty to derive the error on the intrinsic color. For those sources which effective temperatures were directly derived, the T\textsubscript{eff} uncertainty was used, with a minimum value of one spectral sub-type.

\subsection{Comments on the red spectral range}\label{red_range}

At the INT and NTT telescopes additional observations at a higher resolution were performed for each source, covering a redder spectral range ($\sim5800-7000$ \r{A}). The exceptions to this are VOS 50 and VOS 4463 (i.e. 104 sources in total, see Table \ref{Table_appexA_1}). This red range covers the important diagnostic H\textalpha{} line, which enables the determination of accretion rates. An example of the normalized spectra obtained in this region around the H\textalpha{} line is shown in Fig. \ref{fig:Example spectra}.

The measured H\textalpha{} equivalent widths (EW\textsubscript{obs}, observed above the continuum) are tabulated in Table \ref{Table_accre}. The H\textalpha{} line profile was classified into single-peaked, double-peaked, or P-Cygni profile (regular or inverse), following the classification scheme of \citet{2018A&A...620A.128V}. In addition, in Table \ref{Table_accre} we state whether the H\textbeta{} line covered in the blue range (Sect. \ref{S_blue_range}) appears in emission.

%In addition, next to H\textalpha{}, the red spectra cover an additional number of lines that are useful for diagnostics of (circumstellar) activity, such as the He 5876 line and the Na D doublet at 5890 \r{A}. It also contains the strong Diffuse Interstellar Bands (DIBs) at 5797 and 5870 \r{A} that trace interstellar extinction.

\section{Stellar parameters}\label{S_stellar_parameters}

In this section we use the Gaia Early Data Release 3 (EDR3, \citealp{2016A&A...595A...1G, 2021A&A...649A...1G}) to derive the stellar luminosity of the observed sources and place them in the Hertzsprung-Russell (HR) diagram. The HR diagram in combination with theoretical tracks provide us with estimations of the sources' stellar mass and age.

\subsection{Data acquisition and calibration}\label{s_data_acq}

We obtained EDR3 source identifications by using the DR2 source identifications provided in \citet{2020A&A...638A..21V} and the \textit{gaiaedr3.dr2\_neighbourhood} table of the Gaia Archive (see \citealp{2021A&A...649A..10T}).

The EDR3 parallax (\citealp{2021A&A...649A...2L}) to distance conversion was done as follows. For the 140 sources with $\sigma(\varpi)/\varpi\leq0.1$ we obtained the distance by simple inversion of the parallax. For the five sources with $0.1<\sigma(\varpi)/\varpi\leq1$ we used the geometric prior of \citet{2021AJ....161..147B}. In all cases the correction to the zero point parallax bias of \citet{2021A&A...649A...4L} was applied. To trace problematic parallaxes that may lead to spurious or heavily inaccurate distances, we used the \textit{Fidelity} parameter of \citet{2022MNRAS.510.2597R}. Only three observed sources have fidelities below $90\%$ (VOS 1385, VOS 1440, and VOS 2158).

%RAVE Madera only as recommended RAVE DR6 paper.

%We can say that the intrinsic colors matches those teff vs color obtained from spectra only with GALAH (Casagrande2020) and RAVE (myself, is in a TOPCAT folder) mention the sigma of the fit?

%Casagrande only intrinsic colors up to 9000K. mention?

%Here we need to extend this and we use the tycho spectral type (ref Wright), including those o Fabricius 2002.. Match at low+RAVE?

%which we assume can be extended to B type stars.

%Although the Gaia EDR3 photometry has improved significantly from DR2 (e.g, \citealp{2021A&A...649A...5F}, \citealp{2021ApJ...908L..24Y}), still

%FINAL. Juts adpat PM2013 to Gaia colors with Riello transformation (from V-Ic to Bp-Rp). Good match to what was done in my thesis, sigma is around 0.049. 

%Some corrections to the Gaia EDR3 photometry are necessary (e.g \citealp{2021A&A...649A...5F}; \citealp{2021ApJ...908L..24Y}). We corrected the Gaia G band magnitude for sources with 6-parameter astrometric solutions following the recommendations of \citet{2021A&A...649A...3R}, where the Gaia photometry is described. 
The Gaia photometry of the sources (described in \citealp{2021A&A...649A...3R}) is presented in Table \ref{Table_log_observations}. To obtain the Gaia EDR3 intrinsic G\textsubscript{BP}-G\textsubscript{RP} colors of the observed sources, we converted the intrinsic Johnson-Cousins colors for dwarf stars of \citet{2013ApJS..208....9P} to Gaia G\textsubscript{BP}-G\textsubscript{RP} colors with the polynomial equation presented in Table C.2 of \citet{2021A&A...649A...3R}. In order to evaluate the validity of this approach, we compared the intrinsic colors derived this way with those obtained from GALAH+ spectra (\citealp[see their Figure 3]{2021MNRAS.507.2684C}), from RAVE DR6 data (\citealp{2020AJ....160...83S}) and from the Tycho-2 Spectral Type Catalogue (\citealp{2003AJ....125..359W}). The correspondence in all cases is within the 0.03 mag error. 

%and the flux excess factor (\textit{phot\_bp\_rp\_excess\_factor})

%We used those extinctions, the distances, and the Gaia G magnitudes to derive absolute magnitudes ($M_{G}$).

%We use the flux excess factor to mark to sources with a possible incorrect color (as an indicator, although this does not necessarily implies that the data was affected by processing problems of variability), and then derive the reddening and extinction by intrinsic colors

%For the sources which spectral types were determined I took one spectral sub-type as uncertainty to derive the error on the intrinsic color. For those sources which effective temperatures were directly derived, the T\textsubscript{eff} uncertainty was used, with a minimum value of one spectral sub-type (using the \citealp{2013ApJS..208....9P} values). Then, extinctions were derived using the Gaia observed colors and the color dependant extinction coefficients of \citet[see their Figure 1]{2021MNRAS.507.2684C}. Finally, I used those extinctions and the Gaia G magnitude and distance to obtain absolute magnitudes. 

Extinctions were obtained using the effective temperatures (derived in Sect. \ref{S_blue_range}), the intrinsic colors, and the color dependent extinction coefficients of \citet[see their Figure 1]{2021MNRAS.507.2684C}. In all cases $R_{V}=3.1$ was assumed. The median error for the derived A\textsubscript{V} values is $0.17$ mag. 

By fitting an atmosphere model from \citet{2004astro.ph..5087C} of the corresponding T\textsubscript{eff} to the dereddened Gaia G\textsubscript{RP} photometry we derive the total stellar flux for each source. Combining this flux with the distance we obtain the total luminosity ($L$, in a procedure similar to that of \citealp{2018A&A...620A.128V}). We have assumed solar metallicity and $\text{log(g)}=4.0$. The effect of these parameters in the derived luminosities is negligible. 

The luminosities of all sources are presented in Table \ref{Table_stellar_param}, together with their distances and  A\textsubscript{V} extinctions.

%We fit to the $G_{RP}$ as it is the one with less reddening. So we use Ar. Used because G white is too broad and includes Balmer emission and so it is G_BP. G also is harder to calibrate the extinction.

%We note that different studies have suggested that the EDR3 parallax uncertainties are slightly underestimated, specially for bright objects (up to $20-30\%$ for sources with $9\lesssim G\lesssim13$ mag, see e.g. \citealp{2021A&A...649A...5F, 2021arXiv210105282E, 2021arXiv210110206M, 2021arXiv210107252Z}). Therefore, the provided distance uncertainties in Tables \ref{Table_stellar_param} and \ref{Table_stellar_param_cont} may be slightly underestimated. 

%Error on G\textsubscript{RP} to assess variability.

%\begin{figure*}[t!]
%\includegraphics[scale=0.647]{HR_diagram_edited_2.pdf}
%\caption{\textit{Left:} HR diagram of the 123 stars observed after removing the 22 contaminants of Sect. \ref{S_contaminants}. The sources with a dubious classification are discussed in Sect. \ref{S_Dubious}. \textit{Right}: The 209 previously known Herbig Ae/Be stars considered in \citet{2021A&A...650A.182G} are added to the left plot. PARSEC 1.2S PMS tracks  and isochrones corresponding to $1$, $2$, $4$, $6$, $8$, $10$, $15$ M$_{\odot}$ and $1$ and $3$ Myr are presented in both panels (\citealp{2012MNRAS.427..127B} and \citealp{2017ApJ...835...77M}).}\label{Plot: HR_diagram}
%\end{figure*} 

\begin{figure}[t!]
\includegraphics[width=\columnwidth]{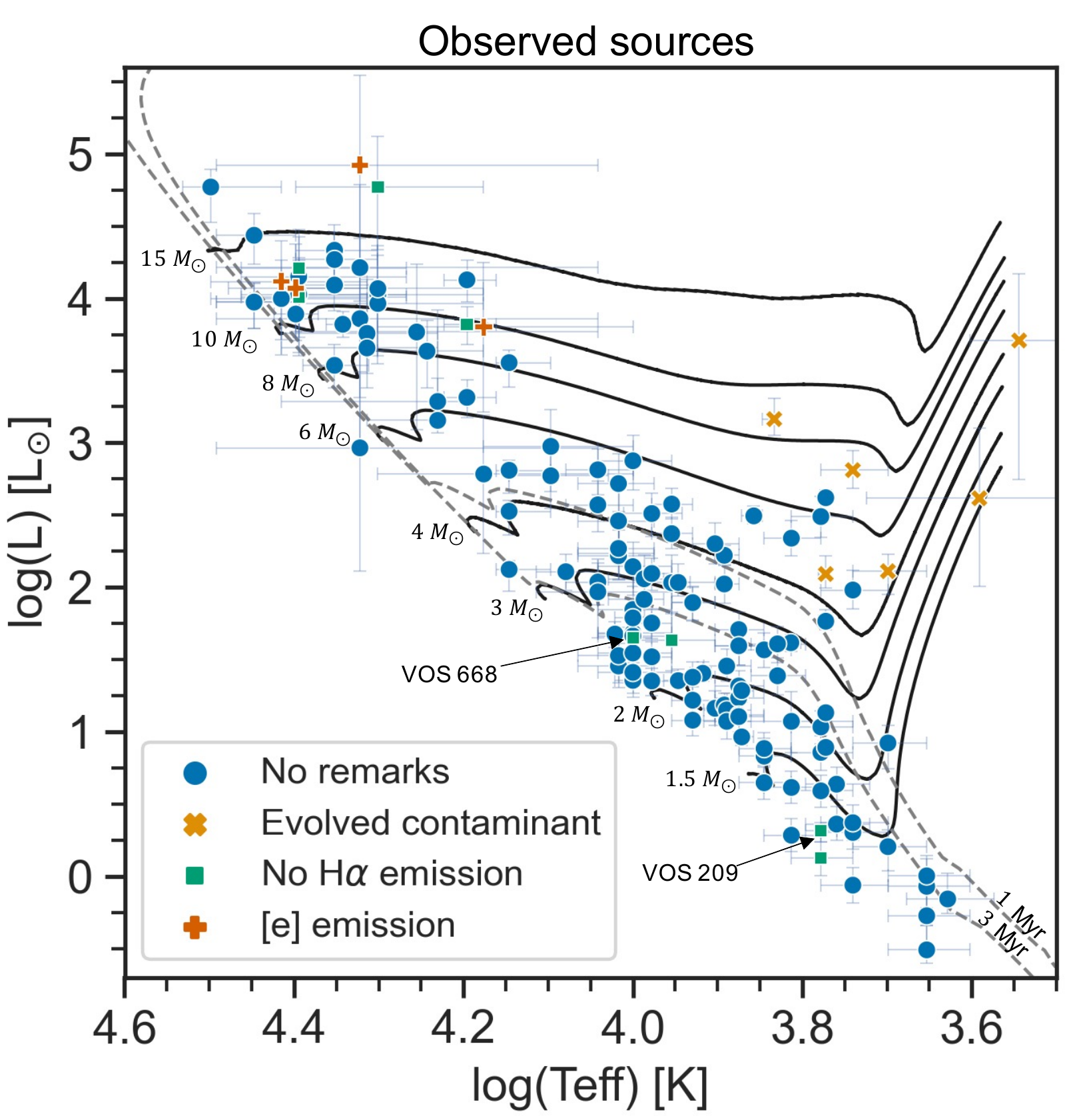}
\caption{HR diagram of the 143 stars observed with T\textsubscript{eff} determinations. The sources with a remark are discussed in Sect. \ref{S_contaminants} and Appendix \ref{S_dubious}. PARSEC 1.2S PMS tracks  and isochrones corresponding to $1.5$, $2$, $3$, $4$, $6$, $8$, $10$, $15$ M$_{\odot}$ and $1$ and $3$ Myr are presented (\citealp{2012MNRAS.427..127B} and \citealp{2017ApJ...835...77M}).}\label{Plot: HR_diagram}
\end{figure}

\subsection{SEDs and IR excess}\label{S_IR_excess}

Fitting an atmosphere model to the dereddened photometry (Sect. \ref{s_data_acq}) allows us to generate Spectral Energy Distributions (SED) to estimate the amount of infrared (IR) excess. For doing this we use the 2MASS and WISE passbands (from AllWISE, \citealp{2013wise.rept....1C}) which are available for all sources (see \citealp{2020A&A...638A..21V}). These passbands range from $J$ band ($1.24$ \textmu{}m) to $W4$ band ($22$ \textmu{}m). The derived IR excesses (L\textsubscript{IR}/L\textsubscript{*}) appear tabulated in Table \ref{Table_accre}.

We note that WISE bands, especially W3 ($12$ \textmu{}m) and W4, have a Point Spread Function that might lead to contaminated photometry in crowded regions or with bright backgrounds (see e.g. \citealp{2014ApJ...791..131K}, \citealp{2014A&A...561A..54R}). In general, the IR excesses derived this way should be considered as indicative, as a fraction of them might be affected by this caveat. A warning flag was included in \citet{2020A&A...638A..21V} catalogue indicating potential problematic W3 and W4 passbands. In Table \ref{Table_accre} this warning flag has been refined by examining the images at W1, W2, W3, and W4 as a set. In total, we found 29 sources where contamination is suspected.

There are six sources (VOS 76, VOS 78, VOS 491, VOS 1240, VOS 1385, and VOS 2161) with an IR excess over 70\% of the stellar bolometric luminosity (L\textsubscript{IR}/L\textsubscript{*}$>0.7$), which is the maximum excess typically observed in Herbig stars (e.g. \citealp{2016A&A...586A...6P}; \citealp{2018A&A...609L...2B}). This implies that the IR excess luminosity of those sources likely includes a large amount of contamination from extended background emission.

%Hence, we note that their classification as PMS candidates might have been compromised by this spurious IR excess. 

%However, we kept them in the analysis as no evidence was found to support a possible non-PMS nature of these sources.

\subsection{Hertzsprung-Russell diagram}\label{S_HR_diagram}

We present the 143 observed sources with T\textsubscript{eff} determinations in an HR diagram in Fig. \ref{Plot: HR_diagram}. From this HR diagram we derive masses and ages by using the PARSEC 1.2S pre-main sequence tracks (\citealp{2012MNRAS.427..127B, 2017ApJ...835...77M}). These masses and ages are tabulated in Table \ref{Table_stellar_param}. We note that the ages are very model dependent, are based on an arbitrary decision of the age `zero', and are very susceptible to the HR diagram location uncertainties. 

There are nine sources which are inconsistent with masses M\textgreater{}$1.5$ M$_{\odot}$. Therefore, they rather belong to the T Tauri regime. These are highlighted in Table \ref{Table_stellar_param}. The remaining 134 sources are compatible with the Herbig regime (M\textgreater{}$1.5\,\textrm{M}_{\odot}$). 

%We note that the general catalogue of new Herbig stars presented in \citet{2020A&A...638A..21V} is HR diagram independent, and hence this result is a compelling positive assessment of that classification.

%\textbf{Better job at deriving the errors, also for the HR diagram. Derivatives would result in smaller errors than just maximizing the expressions}

%\textbf{Would it be possible to derive accretion rates from other parameters?, from Balmer excess? IGAPS?, it seems complicated and unecessary on a first impression.}

%In addition to the observables collected in Chapter \ref{Chapter2} I performed a 1 arcsec cross-match between these samples and IGAPS (INT Galactic Plane Survey, \citealp{2020arXiv200205157M}). IGAPS is the merger of IPHAS and UVEX (The UV-Excess survey of the northern Galactic Plane, \citealp{2009MNRAS.399..323G}), although it is entirely independent of IPHAS as a catalogue. The main add on is that it provides the U\textsubscript{RGO} passband ($0.36\mu m$), a catalogue of emission line stars and another catalogue of variable stars. The sources with U\textsubscript{RGO} information are 4254/8248 ($\sim52\%$), 527/636 ($\sim83\%$) and 770/1264 ($\sim61\%$) for the PMS, CBe and Either samples respectively.

%- Many objects with emission are missed with the emission indicator of IGAPS, specially the massive  extincted ones (Fig 6 of Drew et al.)

%I still need to derreden to U band to show the plot of Lacc vs. U-r. In Fig below it is not dereddeneded.

\section{Contaminants}\label{S_contaminants}

The spectral analysis of Sect. \ref{s_observations} and the derivations of Sect. \ref{S_stellar_parameters} allow us to identify some evolved contaminants among the observed PMS candidates. VOS 1634 and VOS 1806 have spectral types corresponding to an M and a K star, respectively. Their Gaia EDR3 parallaxes assign them luminosities over 400 L$_{\odot}$. Hence, these are probably post-MS giants rather than PMS sources, as YSOs this massive would not be optically visible at such early stages of evolution. VOS 2164 spectra clearly correspond to a planetary nebula (and it appears as so in previous literature; e.g. \citealp{1976ApJ...203..636K}). In addition, previous literature allowed us to identify VOS 603 as a cataclysmic variable star (dwarf nova; e.g. \citealp{2016MNRAS.460.2526O}), VOS 1240 as a carbon star (e.g. \citealp{2002A&A...390..501G}), VOS 1380 as a Type II Cepheid (e.g. \citealp{2004AJ....128.1748S}), and VOS 1385 as a RV Tau variable (\citealp{2018A&A...619A..51B}). VOS 458 was identified as an AGB candidate in \citet{2008AJ....136.2413R}. However, it is a bit hot and under-luminous for an AGB star. We included VOS 458 in the list of contaminants because of its uncertain nature, although we note that the stellar parameters derived for VOS 458 are not incompatible with a YSO nature. 

%This could be done for all sources but for five PMS candidates. Gaia DR2 4318785810234714752, 506799479443438080 and 5333545642950621696 could not be spectral typed, mainly because of their strong emission line spectrum. Thus, the observed PMS candidates Gaia DR2 431934385541454080 and 2071705173505640448, which are a late-K/M and M1 star respectively, do not have intrinsic colors tabulated in Table \ref{intrinsic_color} and no extinction or absolute magnitude were derived for them. We note that these five objects are very likely contaminants: Gaia DR2 4318785810234714752 is a known planetary nebula (and was tagged with a `PN' flag in my catalogue) and 506799479443438080 appears as dwarf nova in SIMBAD (from e.g. \citealp{2015MNRAS.451.2863S} or \citealp{2018A&A...617A..26D}). Gaia DR2 5333545642950621696 has a PN warning flag in my catalogue and might be a B[e] (FS CMa) star. This is supported by the extremely strong emission it shows in all H lines observed (see Sect. \ref{S_Contaminants}). Gaia DR2 431934385541454080 and 2071705173505640448 are probably evolved stars given their low temperature and low absolute magnitude (high luminosity) in the HR diagrams of Chapter \ref{Chapter2}.

In total, we have identified eight evolved stars within the 145 observed PMS candidates. These appear marked in the HR diagram of Fig. \ref{Plot: HR_diagram} and are removed from the rest of the analysis. Hence, the number of observed sources we consider in what follows is $137$ ($145-8$).

We should point out that separating Herbigs from classical Be stars (very similar emission-line non-PMS sources, \citealp{2013A&ARv..21...69R}) was the main task of the machine learning algorithm used in \citet{2020A&A...638A..21V}. Hence, the sample observed in this work is already filtered of classical Be stars.

\section{Mass accretion rates}\label{S_accre}

%We observed the H\textalpha{} line for 86 sources of the 123 that passed the filter of Sect. \ref{S_contaminants}. 

For the sources with H\textalpha{} and H\textbeta{} lines in emission we corrected the measured EWs for the underlying line absorption. To do this we used the typical EW absorption values of each spectral sub-type (\citealp{2015AJ....150..204J}). These corrected equivalent widths (EW\textsubscript{cor}) are tabulated in Table \ref{Table_accre}, together with the observed ones (EW\textsubscript{obs}). Following, for example, \citet{2017MNRAS.464.4721F}: \(F\textsubscript{H\textalpha{},\textbeta{}} = EW\textsubscript{cor}\cdot F_{\lambda}\), where $F_{\lambda}$ is the continuum flux density corresponding to the central wavelength of the H\textalpha{} or H\textbeta{} line. We obtained this flux by using atmosphere models from \citet{2004astro.ph..5087C}, which were scaled to the dereddened G\textsubscript{RP} flux of each star. Then, \(L\textsubscript{H\textalpha{},\textbeta{}} = 4\pi d^2 \cdot F\textsubscript{H\textalpha{},\textbeta{}}\), where $d$ is the distance to the sources. Following \citet[see also \citealp{2011A&A...535A..99M}, \citealp{2020MNRAS.493..234W}, and references therein]{2017MNRAS.464.4721F} we can derive the accretion luminosity (L\textsubscript{acc}) from the lines as:

%As it is often hard to measure that continuum in H\textalpha{} (\textbf{I don't think this is true, I don't know why we do not have this info}), we used the very close photometric point $r$ from the IPHAS or the VPHAS+ survey (\citealp{2014MNRAS.444.3230B} and \citealp{2014MNRAS.440.2036D} respectively). The effective wavelength of $r$ is at only $414$ \AA{} difference from the H\textalpha{} line.

%Hence:

%\begin{equation}
%   F\textsubscript{H\textalpha{}}\sim EW\textsubscript{cor}\cdot F\textsubscript{r} = EW\textsubscript{cor}\cdot F\textsubscript{r\textsubscript{0}}\cdot 10^{(A_{r}-r)/2.5},
%\end{equation}

%were we derived the extinction at $r$ using the extinction law from \citet{1999PASP..111...63F}.

\begin{equation}
    \text{log}(L\textsubscript{acc}/L_{\odot}) = A + B \cdot\label{Eq_Line_lum} \text{log}(L\textsubscript{H\textalpha{},\textbeta{}}/L_{\odot}),\label{eq_3}
\end{equation}

where A and B are constants. For Herbig stars, \citet{2017MNRAS.464.4721F} determined these constants to be $A=2.09\pm0.06$ and $B=1.00\pm 0.05$ for the H\textalpha{} line and $A=2.60\pm0.09$ and $B=1.24\pm 0.07$ for the H\textbeta{} line.

Finally, the mass accretion rate ($\dot{M}\textsubscript{acc}$) can be derived as:

\begin{equation}
    \dot{M}\textsubscript{acc} = \frac{L\textsubscript{acc}R_{*}}{GM_{*}} = \frac{L\textsubscript{acc}}{GM_{*}}\cdot\sqrt{\frac{L}{4\pi\sigma {T_{\text{eff}}}^{4}}},\label{eq_4}
\end{equation}

using the stellar parameters derived in Sect. \ref{S_stellar_parameters}. In case information about both the H\textalpha{} and H\textbeta{} line is available the derived accretion rates come from the H\textalpha{} line EW\textsubscript{cor} (this was done for consistency, as stars with H\textalpha{} emission often have H\textbeta{} in absorption). We derive accretion rates for 92 sources using the H\textalpha{} line and for 12 other sources using the H\textbeta{} line. The mass accretion rates derived this way are tabulated in Table \ref{Table_accre}. We note that Eqs. \ref{eq_3} and \ref{eq_4} assume a magnetospheric accretion mechanism (\citealp{2016ARA&A..54..135H} and references therein).

\section{PMS nature of the observed sources}\label{S_PMS_nature}

In this section, we assess the PMS nature of the observed sources by means of their location in the HR diagram, IR excesses, line profiles, and accretion rates. We note that it is beyond the capabilities of the data presented in this paper to assert with absolute certainty whether all the observed sources are indeed new Herbig discoveries. In fact, it even proved difficult and controversial for much more intensely studied objects (e.g. HD 45677, \citealp{2017ASPC..508....3O}). Nevertheless, in this section we provide ample and independent evidence to conclude that the vast majority, if not all, of the 137 considered sources ($145-8$, Sect. \ref{S_contaminants}) are of a Herbig nature (see Appendix \ref{S_dubious} for a description on the most dubious sources).

\subsection{HR diagram, stellar masses, and IR excesses}

\begin{figure}[t!]
\includegraphics[width=\columnwidth]{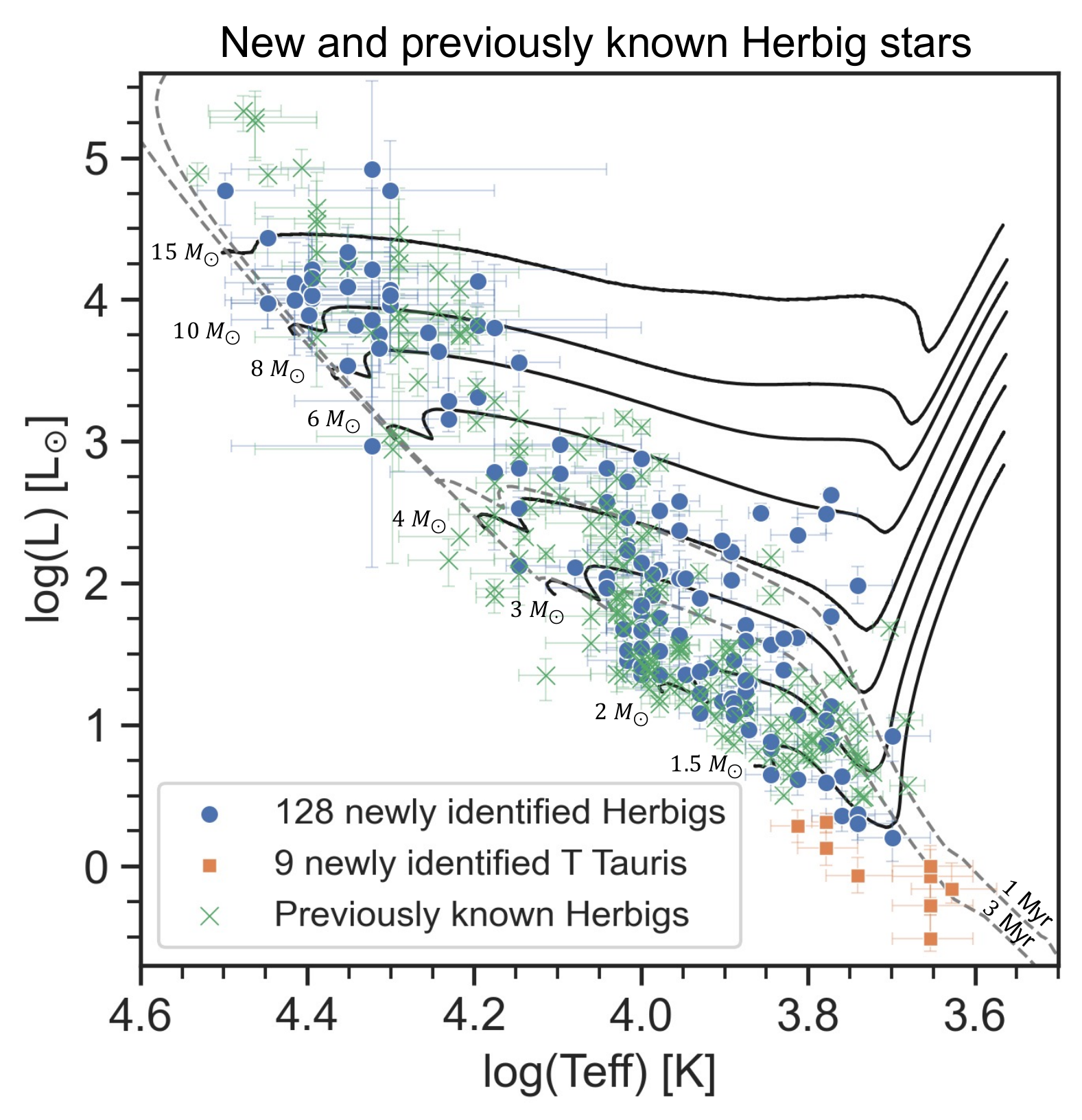}
\caption{HR diagram of the 137 observed sources identified as PMS stars. Previously known Herbig Ae/Be stars and IMTTs with good astrometric quality are also shown. PARSEC 1.2S PMS tracks  and isochrones corresponding to $1.5$, $2$, $3$, $4$, $6$, $8$, $10$, $15$ M$_{\odot}$ and $1$ and $3$ Myr are presented (\citealp{2012MNRAS.427..127B} and \citealp{2017ApJ...835...77M}).}\label{Plot: HR_diagram2}
\end{figure}

In the HR diagram of Fig. \ref{Plot: HR_diagram2} it can be seen that most of the 137 observed sources are massive hot objects. In addition, none of the observed sources are located outside PMS locations in the HR diagram, although the proximity of some of them to the Zero Age Main Sequence (ZAMS) hampers a clear PMS identification. In the HR diagram of Fig. \ref{Plot: HR_diagram2} we also show previously known Herbig Ae/Be stars and IMTTs (from the compilations of \citealp{2018A&A...620A.128V}, \citealp{2021A&A...650A.182G}, and \citealp{2021A&A...652A.133V}) with good astrometric solutions [$\sigma(\varpi)/\varpi\leq1$, $\text{RUWE}<2$, and $\text{Fidelity}>0.95$]. The stellar parameters of these previously known sources were rederived following Sect. \ref{S_stellar_parameters} procedures to compare HR diagrams which are affected by the exact same systematics and uncertainties. We conclude that the observed sources are similarly distributed in the HR diagram to the previously known Herbigs.

\begin{figure}[t!]
\includegraphics[width=\columnwidth]{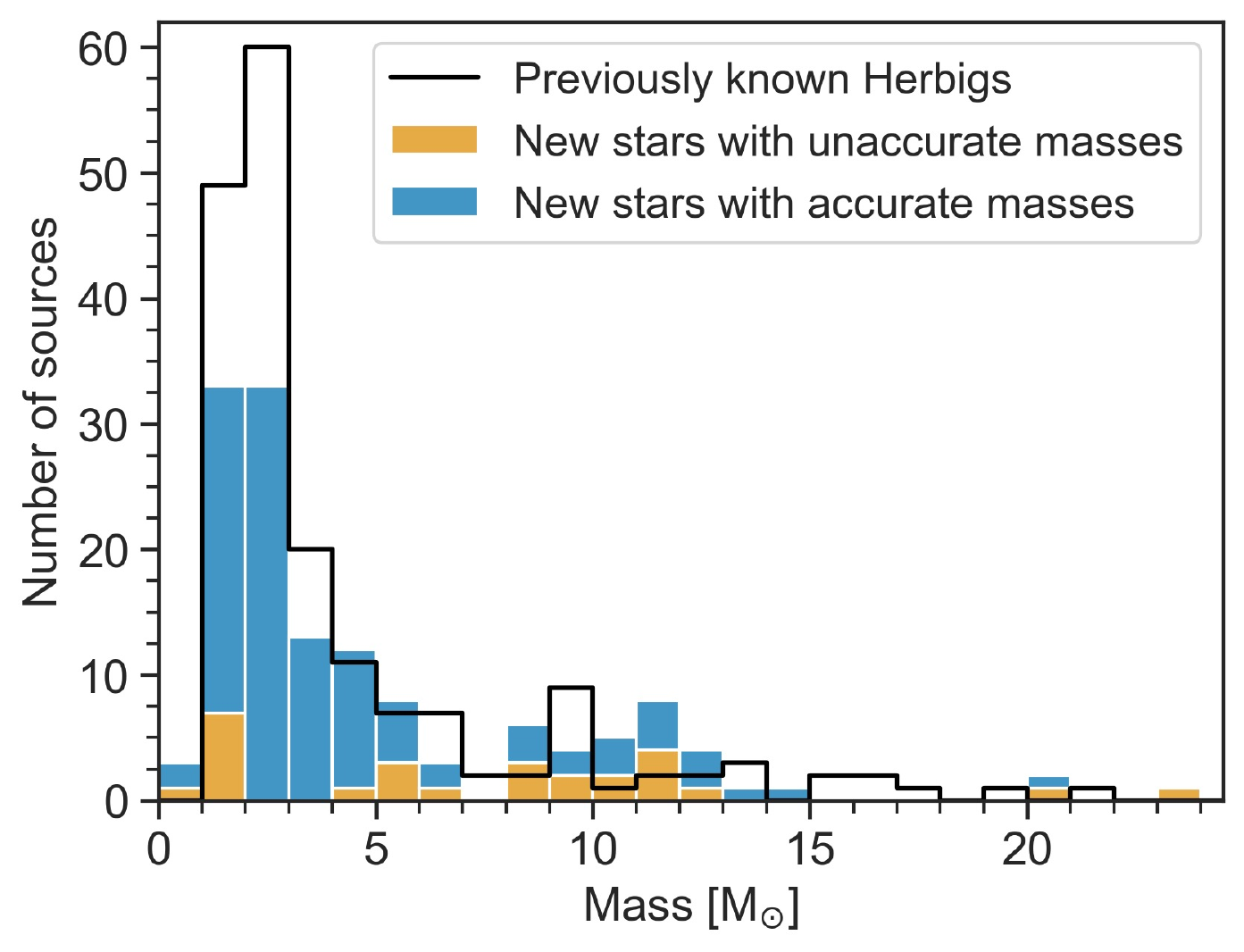}
\caption{Histogram of the number of stars observed in this work per $1$ M$_{\odot}$ bin. In blue the 110 observed stars with stellar mass determinations larger than three times the uncertainty [$M>3\sigma(M)$]. In orange the 27 stars with $M<3\sigma(M)$. Contours trace the previously known Herbig Ae/Be stars and IMTTs with good astrometric quality.}\label{Plot_hist_mass}
\end{figure}

In Fig. \ref{Plot_hist_mass} we present the mass distribution of the 137 sources. In this figure we also show the mass distribution of the previously known Herbig Ae/Be stars and IMTTs. We note that the observed sources and the previously known Herbigs cover a similar mass range and have an analogous mass distribution. This is likely due to the fact that both groups are tracing the massive end of the IMF (see \citealp{2018A&A...620A.128V, 2021A&A...650A.182G}). It is noteworthy that only three sources with M\textgreater{}$15$ M$_{\odot}$ were observed, whereas there are several previously known Herbig stars over that mass. This is because PMS sources with those masses, in addition to being rare and often optically obscured, are typically at large distances and thus tend to have poor parallaxes. This caused those sources to be systematically excluded from the target selection of Sect. \ref{s_observations}, which sampled the catalogue of Herbig candidates of \citet{2020A&A...638A..21V}.

In Fig. \ref{Plot_IR_Excess_vs_mass} we plot the IR excesses derived in Sect. \ref{S_IR_excess} for the observed sources as a function of mass. In this figure we also show the values obtained for the previously known Herbigs in \citet{2018A&A...620A.128V}. The observed sources have similar IR excesses to the previously known Herbigs, and the break in inner disk dispersion efficacy at $7$ M$_{\odot}$ discussed in \citet{2018A&A...620A.128V} is also present for the observed objects. 

%This break at $7$ M$_{\odot}$ is also present using the optical variability indicator of \citet{2018A&A...620A.128V}.

%The higher red resolution allows us to better resolve the line-profiles, which is important for further assessing the nature of the candidates.

\subsection{Emission lines and accretion rates}

Regarding the presence of emission lines, 92 of the 100 sources with H\textalpha{} line observations show the line in emission. In addition, as discussed in Appendix \ref{S_dubious}, of those eight sources without H\textalpha{} emission, seven show other emission lines. Of the 37 sources for which no H\textalpha{} line information is available 12 have emission in H\textbeta{}. We note that H\textbeta{} emission may not be present even if H\textalpha{} emission is. This high percentage of sources with hydrogen emission supports the PMS nature of the group.

%(e.g. see \citealp{2017MNRAS.464.4721F} for an analysis of the emission lines of previously known Herbigs).

Regarding the H\textalpha{} line profiles, of the 92 stars with H\textalpha{} emission, 37 show single-peaked emission, 44 double-peaked emission, and 11 P-Cygni emission (of which five have inverse P-Cygni profiles). Therefore, $40\%$ are single-peaked, $48\%$ are double-peaked and $12\%$ are P-Cygni. These percentages are similar to those observed for known Herbigs ($31\%$, $52\%$, and $17\%$, respectively; see \citealp{2018A&A...620A.128V}). We suspect that the small difference between both groups is caused by the lower resolution of our observations, which moved many P-Cygni and double-peak profiles to the `single-peaked' group. We point out that similar percentages were found for the Br\textgamma{} line by \citet{2022ApJ...926..229G} in known Herbig stars.

%Indeed, the main references for the line profiles of the known HAeBes in \citet{2018A&A...620A.128V} were \citet{2003AJ....126.2971V} and a private communication of the spectra used for \citet{2017MNRAS.464.4721F}. Both studies have a spectral resolution of $R\sim10000$, which is two to five times larger than the resolution of our observations. 

The high fraction of double-peaked line profiles is also suggestive of the PMS nature of the observed sources; stellar activity would not necessarily result in double-peaked emission profiles and stellar winds would have resulted in a larger fraction of P-Cygni like shapes. The fraction of single-peaked H\textalpha{} profiles is consistent with the fraction of pole-on accretion disks that would be expected in a random distribution.

\begin{figure}[t!]
\includegraphics[width=\columnwidth]{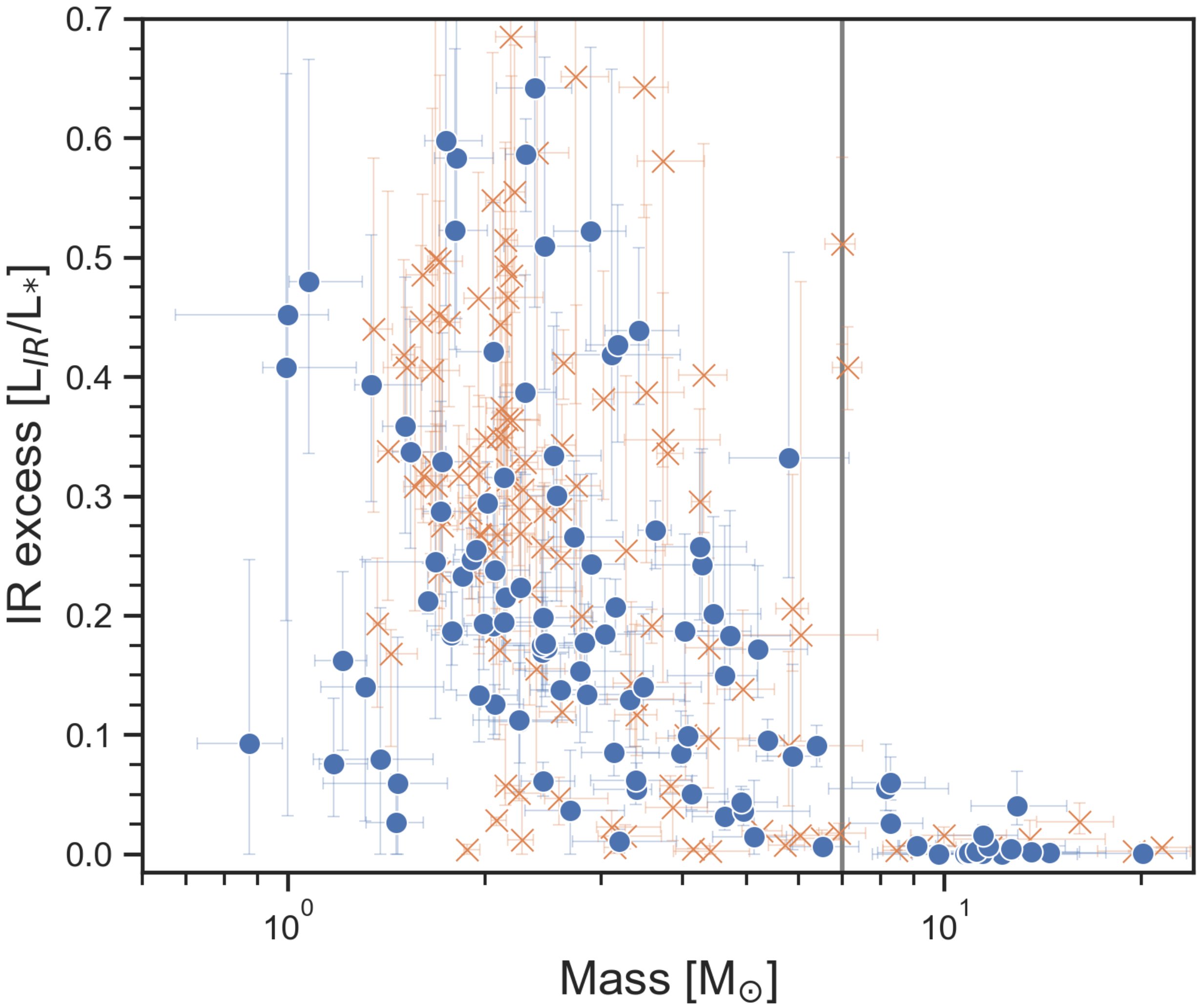}
\caption{Mass vs. IR excess (L\textsubscript{IR}/L\textsubscript{*}). In blue the stars observed in this work with $M>3\sigma(M)$. Orange crosses show the IR excesses derived in \citet{2018A&A...620A.128V} for the previously known Herbigs with a good astrometric solution and $M>3\sigma(M)$. The gray line traces the $7$ M$_{\odot}$ break in inner disk dispersion efficacy.}\label{Plot_IR_Excess_vs_mass}
\end{figure}

\begin{figure}[t!]
\centering\includegraphics[width=\columnwidth]{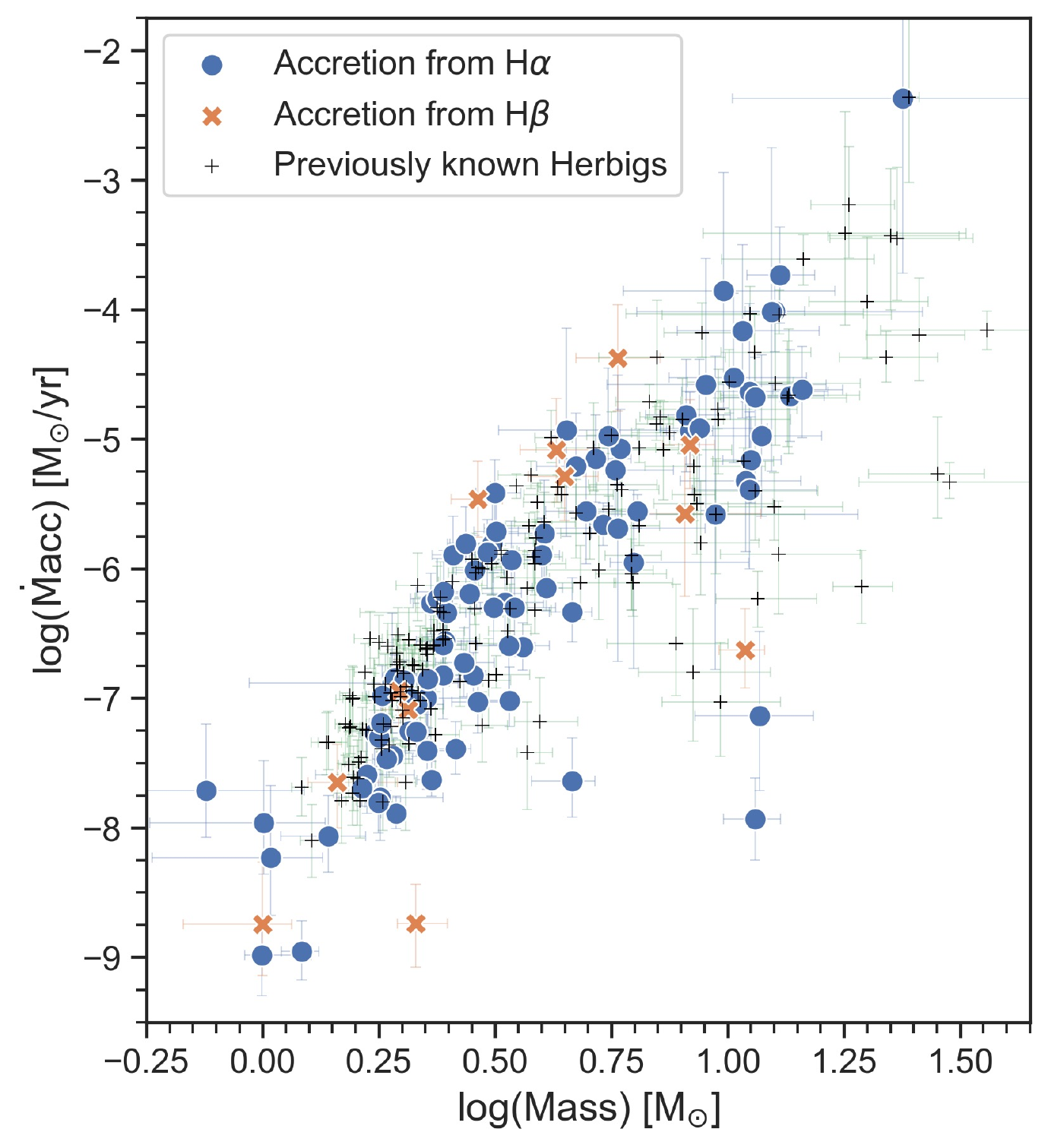}
\caption{Mass vs. mass accretion rate. Blue filled circles indicate the 92 observed stars with accretion rates determined from H\textalpha{} measurements. Orange crosses indicate the 12 observed stars with accretion rates determined from H\textbeta{} measurements. Black plus symbols trace the 163 previously known Herbig stars with mass accretion rates derived in \citet{2020MNRAS.493..234W}.}\label{Plot: accretion}
\end{figure}

The accretion rates of this group of 92 sources with H\textalpha{} emission and 12 sources with H\textbeta{} emission are presented in Fig. \ref{Plot: accretion} as a function of mass. In Fig. \ref{Plot: accretion} we compare these accretion rates with those obtained via the same procedure and assumptions in \citet{2020MNRAS.493..234W} for a set of 163 previously known Herbig Ae/Be stars. The overlap between both sets in this parameter space is consistent with the observed sources being of a Herbig nature. This further supports that the H\textalpha{} and H\textbeta{} emission used to derived the accretion rates is originated in a PMS accretion disk. There are, however, a few outliers in the trend of Fig. \ref{Plot: accretion}. The accretion rates of the observed sources are analyzed in more detail in Sect. \ref{S_accretion_analysis}.

%In Fig. \ref{Plot: accretion} it can be noticed that in the regime M\textless{}2.5 M$_{\odot}$ (log(M)\textless{}0.4) the new sources typically have lower accretion rates than the previously known objects. This is likely due to a selection effect. Previous studies have focused mostly on bright, near-by objects with moderate to strong emission lines, and the near-by sources that were yet to be identified were those with less emission and hence less accretion. In the HR diagram of Fig. \ref{Plot: HR_diagram} we cannot discern any age difference between both groups though, thus hinting at scenario in which accretion does not continuously decrease with time during the PMS phase, but rather does it in episodic outbursts of decreasing intensity (see \citealp{2016ARA&A..54..135H} and references therein).

%In Fig. \ref{Plot_HR_Observed} pre-main sequence tracks of $1$, $2$, $4$, and $7$ M$_{\odot}$ are shown together with the theoretical main sequence described in Sect. \ref{Sec_intrinsic}. 

%In particular, I have found the same correlation of the stellar mass with the H\textalpha{} EW and the photometric variability (Figs. \ref{EWs_mass} and \ref{Plot_var_new}, respectively), and also the same correlation between the H\textalpha{} line profiles and the photometric variability (Fig. \ref{Plot_var_new}).

In this section we have shown evidence for the Herbig nature of most of the observed sources. This can be summarised by their PMS location in the HR diagram (Fig. \ref{Plot: HR_diagram}), the amount of IR excess and its correlation with stellar mass (Fig. \ref{Plot_IR_Excess_vs_mass}), the presence of emission lines in their spectra, the shapes of the H\textalpha{} line, and the derived mass accretion rates (Fig. \ref{Plot: accretion}). Therefore, of the 145 observed objects, we propose 128 sources to be new Herbig identifications, nine sources to be new massive T Tauri identifications (Sect. \ref{S_HR_diagram}), and eight sources to be evolved stars of non-PMS nature (see Sect. \ref{S_contaminants}). A closer look to the 128 proposed Herbig stars allowed us to determine 20 less secure identifications. These 20 stars are discussed in Appendix \ref{S_dubious}.

%These are: VOS 448, VOS 491, VOS 495, VOS 821, VOS 854, VOS 879, VOS 1225, VOS 1276, VOS 1405, VOS 1440, VOS 1771, VOS 1913, VOS 1922, VOS 2051, VOS 2060, VOS 2158, and VOS 2161.

%\textbf{The shaded area indicates the rough region where mass accretion rates are barely measurable because of our EW resolution ($\sim0.15$ \AA{}). We have observed 33 stars without emission. If any of those stars has emission below our detection limit, they would lie in this area.}

\section{Accretion properties}\label{S_accretion_analysis}

The mass accretion rates derived in Sect. \ref{S_accre} from H\textalpha{} and H\textbeta{} luminosities, together with the similar results of \citet{2020MNRAS.493..234W}, allow us to construct the largest sample of Herbig stars with mass accretion rate determinations to date. In total, we compile 258 Herbig sources with derived accretion rates between both works. We note that 48 of these sources have accretion rates that were derived in \citet{2015MNRAS.453..976F} by measuring the UV excess over the Balmer jump. This is a more direct measurement of the accretion rate, and it is free of the assumptions made when correlating accretion luminosity and emission line luminosities (see \citealp{2015MNRAS.452.2837M}).

With this enhanced sample, we revisit the accretion rate properties of Herbig stars. In Fig. \ref{Plot: accretion} we show the accretion rate of this combined sample as a function of stellar mass. We note that for both new and previously known sources the mass accretion rate increases with stellar mass. As \citet{2020MNRAS.493..234W} discussed, we also find that the accretion rate decreases with time during the PMS phase. However, it is not trivial to disentangle this effect from the dependence of the age on the mass.

In \citet{2020MNRAS.493..234W} it was concluded that lower mass Herbigs have a dependence of the mass on the accretion rate, characterised by a gradient that matches the gradient observed in the T Tauri regime (e.g. \citealp{2004AJ....128.1294C,2006A&A...452..245N}). However, higher mass Herbigs show a smaller gradient in the mass vs. accretion rate relation. The break between both groups was set at $3.98^{+1.37}_{-0.94}$ M$_{\odot}$. This accretion break and similar accretion gradients to those found in \citet{2020MNRAS.493..234W} have also been identified by \citet{2022ApJ...926..229G}, using a similar sample and the Br\textgamma{} line as the accretion tracer. 

%This further indicates a difference between the accretion properties of low- and high-mass Herbig stars.

%This was understood as being due to the low-mass objects being subject to magnetospheric accretion (see \citealp{2007prpl.conf..479B}), whereas a fraction of the more massive objects are accreting through a different mechanism; possibly the boundary layer accretion mechanism (\citealp{2020Galax...8...39M}). 

\begin{figure*}[t!]
\centering\includegraphics[width=\textwidth]{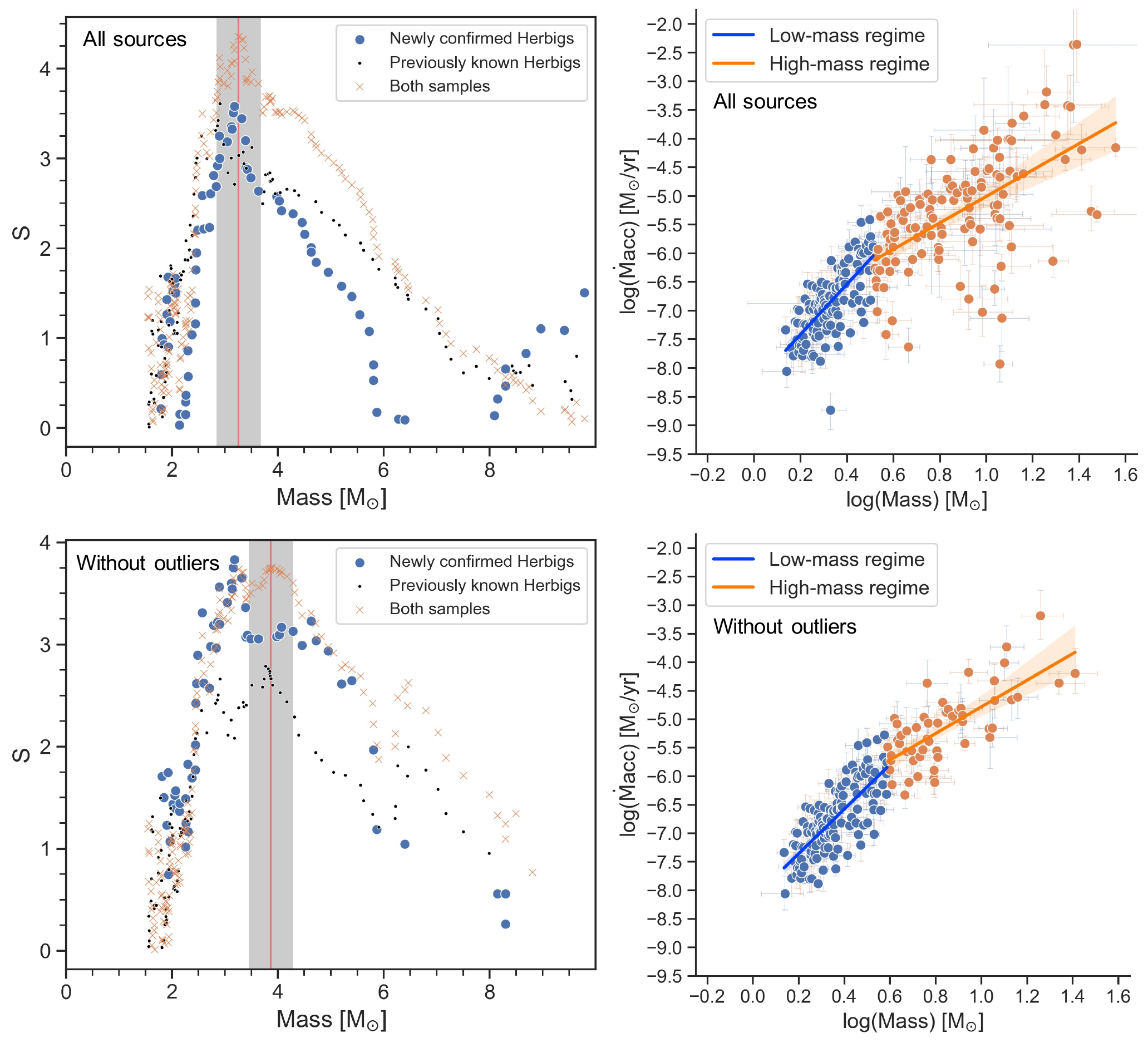}
\caption{\textit{Left plots:} Stellar mass used to separate the low-mass regime from the high-mass regime as a function of the S parameter. Red line traces the maximum S value when all sources are considered, with its estimated uncertainty (gray shaded region). \textit{Right plots:} Mass vs. mass accretion rate for the low- and high-mass regimes, which were defined by the maximum S value obtained in the plots on the left. The fits that gave the gradients with the maximum S value are shown, with a 95\% confidence interval.}\label{PLot_accre_statis}
\end{figure*}

In a similar way to what was done in Section 5 of \citet{2020MNRAS.493..234W}, we looked for the mass value where the difference between the gradients of the low- and high-mass regimes maximizes. We note that this approach does not take into account the uncertainties in the accretion rate, nor the caveats of estimating the accretion rate from emission lines (see e.g. \citealp{2017MNRAS.464.4721F}; \citealp{2020Galax...8...39M}). To study this difference between the observed gradients we use the $S$ parameter. This parameter represents the significance of the difference of the slopes. It is defined as:

\begin{equation}
    S=\frac{|b_{1}-b_{2}|}{\sqrt{\sigma_{b_{1}}^2+\sigma_{b_{2}}^2}},
\end{equation}

where $b$ is the slope in each regime of the mass vs. accretion rate linear fit in log space, and $\sigma^2$ is the variance in that slope. The results of this study on the whole sample of 258 sources are illustrated in the top panels of Fig. \ref{PLot_accre_statis}. In the top left panel we show the S parameter as a function of the mass used to separate the low- from the high-mass regime. We calculated the S parameter for the sample of new Herbigs, the sample of previously known Herbigs, and the combination of those two samples. In all cases the S parameter increases with mass, up to a maximum at around 3-4 M$_{\odot}$, and then monotonically decreases. Using a simple t-test approach we conclude that the difference between gradients is significant to within 95\% confidence if $S\gtrsim2$. Hence, it is evident from Fig. \ref{PLot_accre_statis} that, within the Herbig regime, there is a change in the gradient of the accretion rate as a function of mass (see, however, Sect. \ref{S_problems_acc} for a description of the caveats to consider when using line luminosities as accretion tracers).

%In the top right panel of Fig. \ref{PLot_accre_statis} we show the mass vs. accretion rate plot divided in the low- and high-mass regimes, as defined by 

The maximum S value found for all sources corresponds to a stellar mass for the break in gradient of $3.26$ M$_{\odot}$ (top right panel of Fig. \ref{PLot_accre_statis}). This number is consistent with the value found in \citet{2020MNRAS.493..234W} of $3.98^{+1.37}_{-0.94}$ M$_{\odot}$. However, in the top right panel of Fig. \ref{PLot_accre_statis} it is noticeable that several sources are far from the general correlation. If we remove those `outliers' from the analysis (reducing the sample to 198 sources, bottom panels of Fig. \ref{PLot_accre_statis}) we obtain a maximum S value at $3.87$ M$_{\odot}$, much closer to the central value of $3.98$ derived in \citet{2020MNRAS.493..234W}.

Although there is a range of masses in which the gradient difference as traced by the S parameter is significant; there is a clear peak for S values in all samples at around 3-4 M$_{\odot}$. This peak stands after removing the sources that deviate the most from the correlation. Thus, we confirm that the break in accretion rate detected in \citet{2020MNRAS.493..234W} in the $3$-$4$ M$_{\odot}$ range holds with the sample of new Herbigs. By using the maximum S values obtained for the different Herbig subsamples, we further constrain the potential break in accretion rate to a mass of $3.87^{+0.38}_{-0.96}$ M$_{\odot}$ (corresponding to the mass of a B7-B8 MS star).

\section{Discussion}\label{S_discussion}

In the previous sections we present and discuss 128 new Herbig stars homogeneously selected and observed, for which we provide accurate stellar parameters. In this section we put these sources in context with the historically considered Herbigs, and explain why these new Herbigs provide interesting insights to the intermediate- to high-mass star formation scenario.

\subsection{General remarks}\label{S_General_remarks}

Among the main caveats of previous studies dedicated to intermediate- to high-mass PMS stars is the low number of sources in any given mass or age range (e.g. \citealp{2018A&A...620A.128V, 2021A&A...650A.182G, 2021AJ....162...28V}). The sample of 128 new Herbig stars contains both low-mass objects at the boundary with the T Tauri regime and very-massive PMS objects (see Fig. \ref{Plot: HR_diagram2} and \ref{Plot_hist_mass}). In the mass range of 1.5 to 4 M$_{\odot}$ we present 73 new sources increasing by 42\% the number of known objects. In the mass range of 4 to 8 M$_{\odot}$ we present 23 new sources, increasing by 70\% the number of known objects (55 sources are above the $4$ M$_{\odot}$ threshold of \citealp{2020MNRAS.493..234W} for the break in accretion properties). In the mass range of PMS stars above 7-8 M$_{\odot}$ (typically considered the MYSO regime) we present 32 new sources, increasing by 80\% the number of known objects. This is the threshold of \citet{2018A&A...620A.128V} from which very effective inner disk-dispersal mechanisms are acting. These statistics, summarised in Table \ref{Table_numbers}, use as reference the sample of historically known Herbigs considered in \citet{2018A&A...620A.128V}, \citet{2021A&A...650A.182G}, and \citet{2021A&A...652A.133V} with mass determinations. However, other less famous objects of the class do exist in the literature, like the recently proposed 58 Herbig Ae/Be sources from the LAMOST survey (\citealp{2021RAA....21..288S, 2022ApJS..259...38Z}), the 13 proposed Herbig Ae/Be stars in the Small Magellanic Cloud (\citealp{2019ApJ...878..147K}), or the 77 IMTTs found in the Carina Nebula (\citealp{2021AJ....162..153N}).

%\textbf{In particular, 55 observed sources are above the $4$ M$_{\odot}$ threshold of \citet{2020MNRAS.493..234W} for the break in accretion properties. Of these, 32 are above the $7$ M$_{\odot}$ threshold of \citet{2018A&A...620A.128V} from which very \textbf{efficient} inner disk-dispersal mechanisms are acting.}

The most massive of the newly discovered Herbig stars appear to overlap with the class of Massive Young Stellar Objects (\citealp{2021A&A...648A..62F}; \citealp{2021A&A...654A.109K}). More than 300 such objects with luminosities larger than $5000$ L$_{\odot}$, corresponding to masses larger than around $8-10$ M$_{\odot}$, are listed as MYSO in the RMS catalogue (\citealp{2013ApJS..208...11L}). These massive young stars are infrared bright due to the large amounts of dusty material obscuring them from sight and, as a result, they were  historically assumed to be optically invisible. Intriguingly however, some were already reported to be visible in the optical. For example, the objects PDS 27 and PDS 37 appear in the RMS catalogue (see also \citealp{2019A&A...623L...5K}), but with V band magnitudes of $\sim$13 mag they are optically bright enough to have been recognised as Herbig Be stars (\citealp{2015MNRAS.452.2566A}; \citealp{2018A&A...620A.128V}). Given that Gaia, with its faint magnitude limit of $\sim$20 mag, pushes the definition of optically visible to much fainter magnitudes, the new Gaia-discovered Herbig stars may well bridge the gap between the (optically bright) Herbig Be stars and the optically faint MYSOs. The fact that this project already adds 32 new sources to this mass regime evidences this hypothesis, while preliminary reports show that around 20\% of the previously optically undetected MYSOs may have a Gaia counterpart (Shenton et al. - private communication). Clearly, Gaia plays an important role regarding the most massive young stellar objects.

\begin{table}
\caption{Number of known Herbig stars per stellar mass range.\label{Table_numbers}}
\hskip-1.0cm
\begin{tabular}{cccc}
\hline
Mass range & Previously known &	This work & \% \\
(M$_{\odot}$) &  & & increase\\
\hline
$1.5$ – $4$ & 173 & 73 & 42\% \\
$4$ – $8$ & 33 & 23 & 70\% \\
$>8$ & 40 & 32 & 80\% \\
\hline
Total & 246 & 128\\
\hline
\end{tabular}
\tablecomments{For the previously known objects we only consider the sources discussed in \citet{2018A&A...620A.128V}, \citet{2021A&A...650A.182G}, and \citet{2021A&A...652A.133V} with mass determinations.}
\end{table}

\begin{figure*}[t!]
\centering\includegraphics[width=\textwidth]{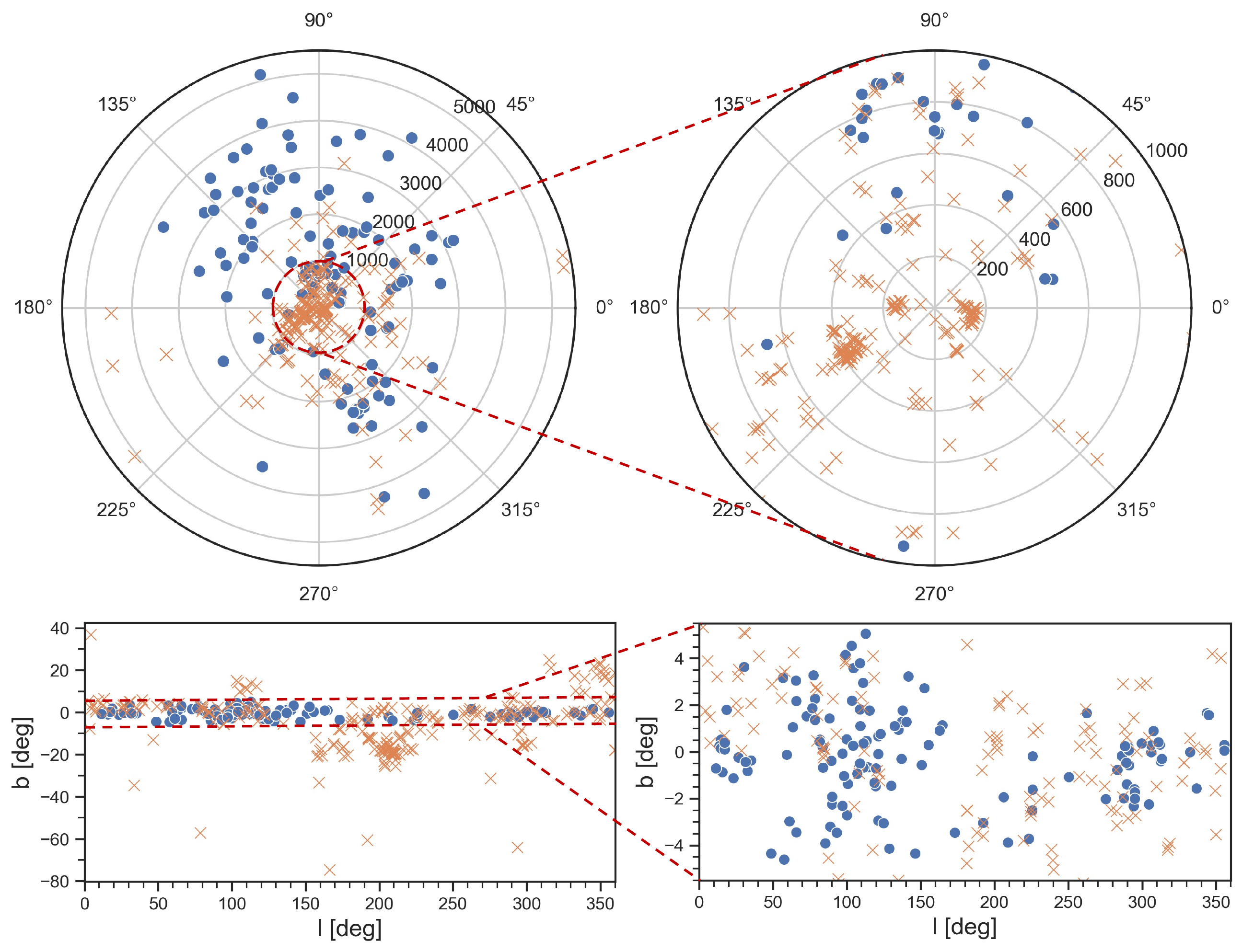}
\caption{The 128 newly confirmed Herbig stars are presented as blue dots. Orange crosses trace all 232 previously known Herbigs in the literature with Gaia EDR3 parallaxes ($\varpi$) and $\sigma(\varpi)/\varpi<1$. Panels on the right are a close up look at the panels on the left. \textit{Top left:} Galactic longitude vs. distance, each circular grid line is 1 kpc, up to 5.5 kpc. \textit{Top right:} Galactic longitude vs. distance, each circular grid line is 200 pc, up to 1000 pc.  \textit{Bottom left:} Galactic longitude vs. Galactic latitude. \textit{Bottom right:} Galactic longitude vs. Galactic latitude limited to $-5.5<b<5.5$ deg, where the newly identified stars are confined.}\label{PLot: Coords_Dist}
\end{figure*} 

%Similarly, the percentage of new MYSOs is affected by the fact that many MYSOs are too optically faint for Gaia and hence they were not included in the aforementioned Gaia-based compilations. However, considering the full sample of 115 well-characterised MYSOs of \citet{2013ApJS..208...11L}, the 32 new MYSOs still suppose an increase of 28\% to number of known MYSOs. 

Regarding the age of the sources, the sample of 128 new Herbig stars contains both sources close to the ZAMS and at earlier stages of evolution. Although some stars could be considered IMTTs (see e.g. \citealp{2021A&A...652A.133V}), most sources clearly belong to the Herbig Ae/Be regime (see Sect. \ref{sec:intro} and footnote 1). With the exception of the four sources discussed in Appendix \ref{S_dubious}, the age range covered at every mass bin is equivalent to the age range covered by previously known and analyzed objects. Given that the \citet{2020A&A...638A..21V} catalogue is HR-diagram independent, this is probably not caused by a selection bias but by the fact that younger objects are too embedded to appear in the Gaia survey.

There are some biases in the catalogue of \citet{2020A&A...638A..21V} that affect the sample of 128 new Herbig stars. These are mainly that sources with strong IR excesses and H\textalpha{} emission were favored in the selection. This biases the sample toward the more `active' PMS objects. We refer the reader to \citet{2020A&A...638A..21V} for more details.

In Fig. \ref{PLot: Coords_Dist} we plot the distances and galactic coordinates of the 128 new Herbig detections. To these we add all previously known Herbigs (Table \ref{Table_numbers}) with Gaia EDR3 parallaxes and $\sigma(\varpi)/\varpi<1$. The new Herbigs are confined to the Galactic plane because of selection effects in \citet{2020A&A...638A..21V}. They are generally fainter than the previously known Herbigs (see Table \ref{Table_log_observations}), and hence they are typically further away. There are only 5 new Herbigs within 500 pc, 80\% of the new sources being beyond 1 kpc.

\subsection{Interpretation of the change in accretion gradient}\label{S_problems_acc}

In this section we interpret the break in accretion gradient at $3-4$ M$_{\odot}$ identified in \citet{2020MNRAS.493..234W} and \citet{2022ApJ...926..229G}, that we extend to the new Herbig stars in Sect. \ref{S_accretion_analysis} and constrain to $3.87^{+0.38}_{-0.96}$ M$_{\odot}$. The most accepted interpretation for this change in accretion gradient is that low-mass objects are subjected to magnetospheric accretion (see \citealp{2007prpl.conf..479B}), whereas a fraction of the more massive objects are accreting through a different mechanism; possibly the boundary layer accretion mechanism (\citealp{2020Galax...8...39M}).

We should caution, however, that the accretion rates derived in Sect. \ref{S_accre} assume a magnetospheric accretion scenario. If a different accretion mechanism applies for some objects their derived accretion rates might be highly inaccurate (see Section 6 of \citealp{2020MNRAS.493..234W}). For example, if a boundary layer accretion mechanism (\citealp{2020Galax...8...39M}) is applying for sources more massive than $3-4$ M$_{\odot}$, the accretion luminosity to accretion rate relation of Eq. \ref{eq_4} would need to be corrected by the relative difference in rotational velocities between the star and the gas contact phase (see Sect. 1.3 of \citealp{2021arXiv211106891W} for more details). In addition, the relation between the line luminosities and the accretion luminosity (Eq. \ref{Eq_Line_lum}) would have to be revisited for the boundary layer scenario. Therefore, we advise the reader to treat the accretion rates derived in this work for massive stars with caution. 

%For example, if a Boundary Layer accretion mechanism (\citealp{2020Galax...8...39M}) is applying for sources more massive than $\sim3.87$ M$_{\odot}$, as it would seem from the break in accretion slope in Fig. \ref{PLot_accre_statis}, the accretion luminosity to accretion rate relation of Eq. \ref{eq_4} would need to be corrected by the relative difference in rotational velocities between the star and the gas contact phase (see Sect. 1.3 of \citealp{2021arXiv211106891W}). If these two velocities differ by a factor of $0.2$, the resulting $\dot{M}\textsubscript{acc}$ would be $\sim3$ times larger than that estimated from Eq. \ref{eq_4}. In addition, the relation between the line luminosities and the accretion luminosity (Eq. \ref{Eq_Line_lum}) would have to be revisited for the Boundary Layer scenario. Therefore, we advise the reader to treat the accretion rates derived in this work for massive stars with caution (Table \ref{Table_accre}).

In addition, there are alternative explanations to the change of trend in hydrogen line luminosity properties, apart from a change in accretion mode (see \citealp{2015MNRAS.452.2837M,2021A&A...652A..68M}). For example, outflowing material could be dominating the line emission (caused by e.g. disk photoevaporation, \citealp{2021A&A...650A.182G}). Indeed, it is not straightforward to characterize the origin of the hydrogen emission lines in high-mass Herbig stars (\citealp[and references therein]{2017MNRAS.464.1984M, 2020Galax...8...39M}). This contrasts to what we observe for the lower mass objects, where the hydrogen emission lines seem to originate in the magnetosphere (\citealp{2020A&A...636A.108B,2020Natur.584..547G}). 

%In this regard, \citet{2021A&A...650A.182G} found that disk photoevaporation is much larger in Herbig Be stars, which might be indicative of a higher incidence of outflowing material.

%\citet{2021A&A...652A..68M} saw that the change of slope in the Rin vs L* diagram from interferometry cannot be explained from a change of accretion mode

%Indeed, it is not certain where the line emission comes from in Herbig stars (see \citealp{2020Galax...8...39M} and references therein).

Furthermore, there is an observational bias to consider. High-mass PMS stars are optically visible for a much shorter time than low-mass PMS stars, and they are often heavily obscured, especially at the younger ages. Hence, in our optical analysis we might be biased against the strongest accretors in the high-mass regime. 

%On the other hand, there is a limit for the lowest accretion rates that we can detect. 

Because of the reasons stated in this section, it is not straightforward to deduce a break in accretion properties only from the break in the behaviour of the hydrogen lines. However, this break at $3-4$ M$_{\odot}$ has been independently detected using different techniques (e.g. near-IR interferometry: \citealp{2005ApJ...624..832M}; optical- and near-UV spectropolarimetry: \citealp{2017MNRAS.472..854A}; spectro-photometry: \citealp{2011A&A...535A..99M}, \citealp{2015MNRAS.453..976F}). It is the combination of those results together with the accretion rates derived from emission lines that lead us to conclude that the break in accretion properties of Fig. \ref{PLot_accre_statis} is likely due to a change between accretion mechanisms; from magnetospheric accretion applying to the lower-mass objects to a boundary layer-like accretion mechanism acting in some, or most, of the more massive objects.

\subsection{Evaluation of \citet{2020A&A...638A..21V} results}

The target list of the observations presented in this work was extracted from the catalogue of new intermediate- to high-mass PMS candidates of \citet{2020A&A...638A..21V}. In this section we reevaluate the accuracy and quality of that catalogue.

In the \citet{2020A&A...638A..21V} catalogue 2226 new Herbig candidates were presented (with $\sigma(\varpi)/\varpi\leq0.2$, this number remains similar when the astrometry is updated to EDR3). We note that, although the catalogue is astrometry-independent, arbitrary cuts to the astrometric quality are necessary to select massive objects with a certain degree of confidence. In this work we have observed 145 objects, which is roughly 6\% of the whole catalogue. As mentioned in Sect. \ref{s_observations}, the target selection was based only on the absolute magnitude and on the parallax quality of the sources. Hence, the target list is representative of the whole catalogue of new intermediate- to high-mass PMS stars of \citet{2020A&A...638A..21V}.

The number of contaminants we have found in this work (8/145 or $5.5\%$) is consistent with the estimated lower-limit precision of the \citet{2020A&A...638A..21V} catalogue ($P\gtrsim81\%$). This affirmation holds true even when we consider as contaminants the more dubious sources of Appendix \ref{S_dubious} (28/145 or $19\%$). We note that the \citet{2020A&A...638A..21V} catalogue is HR diagram independent and thus the high proportion of massive stars targeted (134/145 sources, $92\%$, are above $1.5$ M$_{\odot}$, see Sect. \ref{S_HR_diagram}) is a compelling positive assessment of that catalogue. Therefore, we conclude that the observations presented in this work give independent support to the quality and robustness of the \citet{2020A&A...638A..21V} catalogue.

In addition, 14 classical Be candidates from the \citet{2020A&A...638A..21V} catalogue were observed (these are non-PMS stars which are typical contaminants in Herbig samples, see \citealp{2020PhDT.........4V}). None of these classical Be candidates could be identified as a misclassification (e.g. by showing a PMS nature). The discussion on the observations of these sources will be presented in an independent paper. 

%These were selected only based on their high classical Be probability values.

%, which was the main goal of Chapter \ref{Chapter2}, but they are not sensitive to other types of contaminants.

Because of sensitivity limitations, mostly candidates at the bright end of the \citet{2020A&A...638A..21V} catalogue were observed (90\% of the observed sources are in the $12$\textless{}G\textless{}$14$ mag range). One could therefore argue that the observed sample is biased. However, the \citet{2020A&A...638A..21V} catalogue is distance independent. Hence, the conclusions that arise from these observations can be extrapolated to the fainter objects of the catalogue, given that they were all selected homogeneously by the machine learning algorithm. 

%We note that it is at the fainter end of the catalogue where the vast majority of the sources are totally novel and unexplored. 

%Finally, in Sect. \ref{SIMBAD_set} I presented the SIMBAD set, a set of sources that were not observed by myself but which have spectral types in the literature and could be placed in the HR diagram of Fig. \ref{Plot_SIMBAD}. Therefore, this is a random subsample of the catalogues of Chapter \ref{Chapter2}, which in general has a greater degree of uncertainties and biases than the subsample of sources selected for my own observations. However, in Sect. \ref{Obs_stellar_parameters} I discussed that most of the PMS and CBe candidates of this SIMBAD set seem to be of a Herbig Ae/Be and CBe nature respectively, according to their positions in the HR diagram. Moreover, the HR diagram of the SIMBAD set is similar to the HR diagram of the observed candidates (Figs. \ref{Plot_HR_Observed} and \ref{Plot_SIMBAD} respectively). This is important because it evidences that the conclusions of this section are not biased by the decisions made when constructing the target list (which, for example, selected some of the best candidates; see Sect. \ref{Sect__our_observations}).

%Thus, some of the observed objects have been already characterised and studied by other surveys and authors. Comparisons with previous literature determinations are done in Sect. \ref{S_contaminants}. 

\section{Conclusions}\label{S_Conclusions}

In this work we discuss the results of the spectroscopic observations of a sample of 145 Herbig candidates from the catalogue of \citet{2020A&A...638A..21V}. The main results and conclusions of these observations are the following:

\begin{itemize}
\item{We propose 128 sources as new `Herbig' identifications (i.e. PMS stars with $M>1.5$ M$_{\odot}$). We provide ample evidence supporting this classification. This evidence is based on their PMS location in the HR diagram, the amount of IR excess and its correlation with stellar mass, the presence of emission lines in their spectra, the shapes of the H\textalpha{} line, and the derived mass accretion rates. Only 5 sources lie within 500 pc, whereas 75\% of the stars are between 1 and 4 kpc. Twenty sources were flagged as less secure identifications.}

\item{We derive extinctions and accurate stellar parameters for all sources, placing them in the HR diagram by means of Gaia EDR3 parallaxes.}

\item{This sample of 128 new Herbig stars increases the number of known objects of the class by $\sim50\%$. The sources are distributed over a representative range in mass and age when compared to previously known Herbig stars. According to classical definitions, most of the observed sources fall within the Herbig Ae/Be or the Massive Young Stellar Objects regime, but some stars can be considered Intermediate-Mass T Tauris. In particular, 23 of the new sources have masses between $4$-$8$ M$_{\odot}$ and 32 sources have masses $>8$ M$_{\odot}$.}

\item{Four sources were identified as new `unclassified B[e]' (FS CMa) discoveries. Nine other sources were also identified as having a PMS nature, but their masses assign them to the T Tauri regime.}

\item{We derive accretion rates for 104 of the new Herbig stars by using hydrogen emission lines (H\textalpha{} and H\textbeta{} luminosities). This is a 60\% increment to the number of Herbig stars with derived accretion rates. The change in accretion gradient as a function of mass in the 3-4 M$_{\odot}$ range described in \citet{2020MNRAS.493..234W} is also present for the new Herbig stars. This provides further support to a change in accretion mechanism happening withing the Herbig regime. We constrain the mass for this possible change to $3.87^{+0.38}_{-0.96}$ M$_{\odot}$ (the mass of a B7-B8 main sequence star).}

\item{There are four sources (VOS 63, VOS 67, VOS 821, and VOS 1635) of $5$-$6$ M$_{\odot}$ which are younger than previously known PMS stars of that mass.}

%These are ideal candidates for follow-up observations on the study of massive star and disk formation.

\item{The sudden decrease in the amount of near- and mid-IR excess at $\sim7$ M$_{\odot}$ described in \citet{2018A&A...620A.128V} for the historically considered Herbig stars is also present for the new sources. For this group of Herbig stars $\text{M}>7$ M$_{\odot}$ corresponds to sources with $\text{T\textsubscript{eff}}\gtrsim15000$ K. This further supports the idea of very effective inner-disk dispersion mechanisms acting on massive stars, like the disk photoevaporation mechanism proposed in \citet{2021A&A...650A.182G}.}

\item{The observations described in this work provide independent support to the accuracy and high-quality of the catalogue of new intermediate- to high-mass PMS stars presented in \citet{2020A&A...638A..21V}, of which these observations constitute a mere 6\%.}

\end{itemize}

These observations yield a well-defined set of new intermediate- to high-mass PMS stars. Contrary to previous samples, these new Herbig stars were homogeneously identified and observed. Therefore, this set of objects will be the basis for future surveys and follow-up observations dedicated to the Herbig group and their protoplanetary disks. The sample of new Herbig stars presented in this work will be complemented by an X-Shooter Very Large Telescope survey focusing on newly identified intermediate-mass T Tauri stars (Iglesias et al. in prep.). All together, the two surveys will increase by a factor of two the number of known intermediate- to high-mass PMS objects, covering representatively all the stages of the optical evolution of massive forming stars.

%In particular, this set of objects will be the basis for future follow-up observations to analyze their accretion properties and their protoplanetary disks.

%and possible sub-stellar companions

%Hence, this catalogue will be of great use for future studies on the formation of stars in the intermediate- to high-mass regime, which will allow us to study the evolution of intermediate-mass PMS sources from early ages to the main sequence in an unbiased way.

%Mention it is a 6% of that catalogue.

%Covering the high-mass and low-age space of stellar parameters (Mass of $2$-$4.5$ M$_{\odot}$ and Age$<$3 Myr), which remains largely unexplored. These observations will increase the number of observed disks in this space of stellar parameters by a factor of 2.5, providing the means to study the evolution and properties of protoplanetary disks around high-mass PMS stars from early ages (including the earliest stages of PMS evolution under $1$ Myr) to the main sequence in a more complete and uniform way, and to perform consistent comparisons with the low-mass regime. Complementary low-resolution spectroscopic observations of a subsample of bright sources have shown the contamination rate is low and $90\pm3\%$ of the stars in the catalogue are bona fide intermediate-mass pre-main sequence stars (private communication). 

%with ALMA and VLT/SPHERE.

%We have not detected the less evolved Herbig Be stars. 

%The user might want to go to the Either catalogue.

%NEED TO CHECK HR DIAGRAM WITH SOURCES WITH HALPHA ONLY

\begin{acknowledgments}
\section*{Acknowledgments}

The STARRY project has received funding from the European Union's Horizon 2020 research and innovation programme under MSCA ITN-EID grant agreement No 676036. Ignacio Mendigutía is funded by a RyC2019-026992-I grant by MCIN/AEI
/10.13039/501100011033. This work has made use of data from the European Space Agency (ESA) mission {\it Gaia} (\url{https://www.cosmos.esa.int/gaia}), processed by the {\it Gaia} Data Processing and Analysis Consortium (DPAC, \url{https://www.cosmos.esa.int/web/gaia/dpac/consortium}). Funding for the DPAC has been provided by national institutions, in particular the institutions participating in the {\it Gaia} Multilateral Agreement. This research has made use of \texttt{IRAF} which is distributed by the National Optical Astronomy Observatory, which is operated by the Association of Universities for Research in Astronomy (AURA) under a cooperative agreement with the National Science Foundation. This research has made use of the TOPCAT tool (\citealp{2005ASPC..347...29T}). In addition, this research used the VizieR catalogue access tool, and the SIMBAD database developed and operated at CDS, Strasbourg, France.

This article is based on observations made in the Observatorios de Canarias del IAC with the Isaac Newton Telescope (INT) operated on the island of La Palma by the Isaac Newton Group of Telescopes in the Observatorio del Roque de los Muchachos. In addition, this article is based on observations collected at the Centro Astronómico Hispano-Alemán (CAHA) at Calar Alto, operated jointly by Junta de Andalucía and Consejo Superior de Investigaciones Científicas (IAA-CSIC). Finally, this article also used observations collected at the European Organisation for Astronomical Research in the Southern Hemisphere under ESO programmes 0104.C-0937(A) and 0104.C-0937(B).

\end{acknowledgments}

\vspace{20pt}
\appendix
\section{Less secure identifications}\label{S_dubious}

Among the 128 new Herbig stars proposed there are 20 sources which, for different reasons, have a less secure PMS nature than the rest. These sources are discussed in detail in this appendix.

Eight sources have H\textalpha{} and H\textbeta{} fully in absorption (VOS 209, VOS 448, VOS 491, VOS 495, VOS 668, VOS 854, VOS 1922, and VOS 2060). However, three of these sources (VOS 448, VOS 495, and VOS 1922) show an asymmetric H\textalpha{} absorption profile, which might hint some H\textalpha{} emission. These eight sources are shown in the HR diagram of Fig. \ref{Plot: HR_diagram}. Although emission in hydrogen lines is historically one of the defining properties of Herbig stars, some intermediate-mass PMS stars lack hydrogen emission. Hence, this fact alone is inconclusive for removing these sources from the PMS category. In addition, of the eight sources for which we do not detect clear hydrogen emission, seven show other emission lines (the exception is VOS 854). However; VOS 448, VOS 495, VOS 1922, and VOS 2060 do not show any significant level of IR excess, and the IR excess of VOS 491 is clearly spurious (see Sect. \ref{S_IR_excess}). Because hydrogen emission lines and IR excess are the main indicators of YSO nature, we label these latter five sources plus VOS 854 as less secure identifications. We note that 37 observed sources lack H\textalpha{} information, and hence the aforementioned analysis could not be applied to them. Of these, seven sources have H\textbeta{} in absorption and display little IR excess (L\textsubscript{IR}/L\textsubscript{*}\textless{}$0.03$). These are VOS 821, VOS 879, VOS 1225, VOS 1276, VOS 1771, VOS 1913, and VOS 2051, which are also labeled as less secure identifications.

We now consider the nature of VOS 209 and VOS 668, which have emission lines and IR excess but lack hydrogen emission. \citet{2021A&A...651L..11V} found that the intermediate-mass YSO HD 152384 has all hydrogen lines strictly in absorption, but has refractory lines in emission. This led \citet{2021A&A...651L..11V} to suggest that HD 152384 is at the late stages of the PMS phase and is surrounded by a tenuous circumstellar disk caused by the collision of rocky planets (see also the extreme debris disks described in \citealp{2021ApJ...910...27M}). We pose VOS 209 and VOS 668 as PMS sources of a similar nature to HD 152384. Indeed, the age estimates derived in Sect. \ref{S_HR_diagram} imply that both VOS 209 and VOS 668 are compatible with being close to the main sequence ($7.27^{+0.10}_{-0.19}$ and $6.61^{+0.16}_{-0.09}$ Myr, respectively). These two sources are marked in Fig. \ref{Plot: HR_diagram}. 

%We studied whether some of the sources we observed without clear hydrogen emission might have a similar nature. 

%(39 observed with CAHA 2.2m plus VOS 50 and VOS 4463)

%Of those 41, 14 show H\textbeta{} emission.

In addition, we found four sources (VOS 1405, VOS 1440, VOS 2158, and VOS 2161) that have so many permitted and forbidden emission lines that the underlying photospheric absorption spectrum is hardly visible. They are reminiscent of the `unclassified B[e]' objects (\citealp{1998A&A...340..117L}), which constitute a class that includes both evolved stars and young Herbig stars. However, it is often hard to decide on the evolutionary nature of the objects. For example, the archetypal unclassified B[e] star HD 45677 may or may not be a young star (see \citealp{2017ASPC..508....3O,2022A&A...658A..81H}). Such sources are also referred to as FS CMa objects (\citealp{2007ApJ...667..497M}). We therefore propose VOS 1405, VOS 1440, VOS 2158, and VOS 2161 to be new `unclassified B[e]' (FS CMa) discoveries. The absence of clear photospheric hydrogen absorption lines combined with multiple emission lines led to this FS CMa classification. The nature of these sources is unclear, but they could still be of a YSO nature. These new `unclassified B[e]' discoveries are highlighted in the HR diagram of Fig. \ref{Plot: HR_diagram}.

Finally, there are four sources (VOS 63, VOS 67, VOS 821, and VOS 1635), which are younger than all previously known Herbig stars of a similar mass (5 – 6 M$_{\odot}$, see Fig. \ref{Plot: HR_diagram2}). PMS objects of this high-mass and young age are expected to be quite embedded. Thus, these sources require a closer look at their nature. VOS 821 was already mentioned in this appendix regarding its lack of observed hydrogen emission and IR excess. However, we do not have any reason to suspect of the PMS nature of the other three objects, although a post-MS nature can neither be entirely discarded. It has been proposed that the FU Ori outbursting phenomena, which can cause T Tauri stars to get bluer and more luminous (e.g. \citealp{2017A&A...605A..77V, 2018MNRAS.475.2642K}), might explain the position of PMS sources in this region of the HR diagram. However, should that be the case we would expect to measure large H\textalpha{} EWs signposting high accretion rates, and that is not the case for these sources. 

%We decided to keep these four sources in the list of confirmed Herbig detections. 

%Pending final confirmation, they might be the first massive YSOs of their age to be observed. Hence, they might constitute the perfect example to study stellar and disk evolution.

The 20 sources discussed in this appendix are flagged in Tables \ref{Table_stellar_param} and \ref{Table_accre}.

\section{Tables}
This appendix contains Tables \ref{Table_log_observations}, \ref{Table_stellar_param}, and \ref{Table_accre}.

\restartappendixnumbering
\renewcommand\theHtable{Appendix.\thetable}

{\catcode`\&=11
\gdef\2020AanA...638A..21V{\cite{2020A&A...638A..21V}}}

%\movetabledown=35mm
\begin{longrotatetable}
%\centerwidetable
\begin{deluxetable*}{lllllcccclccc}
\tabletypesize{\scriptsize}
%\tablewidth{10 pt}
\tablecolumns{13}
\tablecaption{Log of observations. Column 1 contains the name of the sources as assigned in this work. Column 2 contains the Gaia EDR3 source id of each source. Column 3 shows other literature names. Columns 4 and 5 are Gaia EDR3 right ascensions and declinations. Column 6 and 7 list the observation dates (for the blue and red settings, see Sect. \ref{s_observations}). Columns 8 and 9 present the exposure times. Column 10 states the spectrograph used in each case. Columns 11 to 13 show the Gaia magnitudes. See Table \ref{Table_appexA_1} for details on the instrumental setups used at each telescope. \label{Table_log_observations}}
\tablehead{
\colhead{Name} & \colhead{Gaia EDR3} & \colhead{Alternative} & \colhead{RA} & \colhead{DEC}  & \multicolumn{2}{c}{Obs Date} & \multicolumn{2}{c}{Exposure Time} & \colhead{Spectrograph} & \colhead{G\textsubscript{BP}}  & \colhead{G}  & \colhead{G\textsubscript{RP}}\\
\colhead{} & \colhead{source id} & \colhead{name} & \colhead{hh:mm:ss} & \colhead{dd:mm:ss.s}  & \colhead{Blue}  & \colhead{Red}  & \colhead{Blue [s]} & \colhead{Red [s]} & \colhead{} & \colhead{[mag]} & \colhead{[mag]} & \colhead{[mag]}}
%\rotate
\startdata
VOS 4 & 5524619735681532160 &  & 08:51:11.43 & –41:38:49.9 & 2020-03-12 & 2020-03-20 & 900 & 1100 & EFOSC2 & 13.45 & 13.20 & 12.64 \\
VOS 7 & 2013474586693428096 &  & 22:54:03.79 & +58:54:01.5 & 2019-09-04 &  & 450 &  & CAFOS & 13.59 & 12.82 & 12.22 \\
VOS 15 & 4054609866596686336 & V731 Sco & 17:33:20.78 & –32:34:25.7 & 2020-03-13 & 2020-03-20 & 1000 & 2000 & EFOSC2 & 14.02 & 13.62 & 12.99 \\
VOS 20 & 2206875422143048960 &  & 22:50:57.07 & +61:23:37.7 & 2019-08-06 & 2019-08-12 & 1000 & 1800 & IDS & 14.99 & 14.22 & 13.37 \\
VOS 22 & 2164505844663760768 &  & 21:11:19.06 & +47:38:47.6 & 2019-08-06 & 2019-08-10 & 600 & 900 & IDS & 12.68 & 12.30 & 11.68 \\
VOS 24 & 4267078153499687296 & V868 Aql & 18:56:47.65 & +01:48:51.2 & 2019-08-06 & 2019-08-11 & 750 & 1300 & IDS & 14.00 & 13.15 & 12.22 \\
VOS 26 & 5337719389118815488 & ES Car & 11:10:13.23 & –60:14:03.4 & 2020-03-12 & 2020-03-20 & 1000 & 1600 & EFOSC2 & 13.56 & 13.19 & 12.70 \\
VOS 32 & 425559107889037568 & FY Cas & 00:38:50.08 & +59:54:10.0 & 2019-08-06 & 2019-08-12 & 1200 & 1500 & IDS & 15.60 & 14.58 & 13.56 \\
\enddata
\vspace{-1cm}
%\tablecomments{}
\end{deluxetable*}
\end{longrotatetable}

%\startlongtable
%%\centerwidetable
%\begin{deluxetable*}{c c c c c c c c c c c c}
%\tabletypesize{\footnotesize}
%\tablewidth{0pt}
%\tablecolumns{15}
%\tablecaption{ordered by right ascension.\label{Table_obs}}
%\tablehead{\colhead{\bf{Isaac Newton Telescope}} & \colhead{} & %\colhead{} & \colhead{} & \colhead{} & \colhead{} & \colhead{}\\
%\colhead{Name} & \colhead{Gaia EDR3} & \colhead{RA} & \colhead{DEC} & %\colhead{Obs date} & \colhead{Exposure time (s)} & \colhead{SNR} & %\multicolumn{3}{c}{Photometry (mag)}\\
%\cline{7-9}
%\colhead{} & \colhead{J2000?} & \colhead{J2000?} & \colhead{August %2019} & \colhead{R456 grism} & \colhead{SNR} & %\colhead{G\textsubscript{BP}} & \colhead{G} &  %\colhead{G\textsubscript{RP}}}
%%\rotate
%\startdata
%1 & 2 & 3 & 4 & 5 & 6 & 7 & 8 & 9 & 10 & 11\\
%1 & 2 & 3 & 4 & 5 & 6 & 7 & 8 & 9 & 10 & 11\\
%1 & 2 & 3 & 4 & 5 & 6 & 7 & 8 & 9 & 10 & 11\\
%\enddata
%%\vspace{-0.8cm}
%\tablecomments{The blue spectral region considered in the three %observing runs ($\sim3300-5400$ \r{A}) covers the main wavelength %range to determine spectral types. This region is especially useful %for the earlier spectral type stars such as A and B stars, as their %spectrum in the wavelength range beyond}
%\end{deluxetable*}

%\section{Derived parameters of the observed sources}

%In this section blah in Table \ref{Table_stellar_param} and Table \ref{Table_accre}.

%\movetabledown=35mm
%\startlongtable
\begin{longrotatetable}
%\centerwidetable
\begin{deluxetable*}{llcccccccl}
\tabletypesize{\footnotesize}
\tablewidth{0pt}
\tablecolumns{10}
\tablecaption{Stellar parameters, distances and extinctions (A\textsubscript{V}) for the 145 observed sources, ordered by name. Effective temperatures (T\textsubscript{eff}) or spectral types derived from spectra (Sect. \ref{S_blue_range}) appear in boldface. T\textsubscript{eff} values or spectral types not in boldface were approximated from the spectral derivations using the conversions of \citet{2013ApJS..208....9P}. See Sect. \ref{S_stellar_parameters} for a description on the derivation of the stellar parameters, distances and extinctions. Stellar radii were derived following Eq. \ref{eq_4}. \label{Table_stellar_param}}
\tablehead{
\colhead{Name} & \colhead{SpType} & \colhead{T\textsubscript{eff} [K]} & \colhead{Distance [pc]} & \colhead{A\textsubscript{V} [mag]} &  \colhead{log(L [L$_{\odot}$])} & \colhead{Radius [R$_{\odot}$]} & \colhead{Mass [M$_{\odot}$]} & \colhead{Age [Myr]} & \colhead{Remarks}}
%\rotate
%\renewcommand{\arraystretch}{1.5}
\startdata
VOS 4 & A0 & $\bf{10000^{+500}_{-500}}$ & $932^{+19}_{-18}$ & $1.813^{+0.083}_{-0.086}$ & $1.356^{+0.089}_{-0.092}$ & $1.59^{+0.36}_{-0.29}$ & $2.14^{+0.13}_{-0.14}$ & $9^{+28}_{-3}$ &  \\
VOS 7 & F2–F8 & $\bf{6500^{+500}_{-500}}$ & $472.4^{+4.0}_{-3.9}$ & $1.79^{+0.26}_{-0.28}$ & $0.61^{+0.11}_{-0.11}$ & $1.60^{+0.54}_{-0.39}$ & $1.34^{+0.26}_{-0.07}$ & $15^{+35}_{-7}$ &  \\
VOS 15 & A6 & $\bf{8000^{+500}_{-500}}$ & $1119^{+27}_{-26}$ & $1.76^{+0.19}_{-0.22}$ & $1.16^{+0.12}_{-0.12}$ & $1.98^{+0.61}_{-0.46}$ & $1.81^{+0.25}_{-0.14}$ & $8.0^{+4.6}_{-1.5}$ &  \\
VOS 20 & \textbf{A9} & $7440^{+60}_{-240}$ & $813^{+13}_{-13}$ & $2.81^{+0.01}_{-0.10}$ & $0.964^{+0.028}_{-0.059}$ & $1.83^{+0.19}_{-0.15}$ & $1.63^{+0.11}_{-0.05}$ & $10.21^{+0.76}_{-0.49}$ &  \\
VOS 22 & B8 & $\bf{12500^{+2500}_{-2500}}$ & $2409^{+63}_{-60}$ & $2.50^{+0.13}_{-0.26}$ & $2.98^{+0.25}_{-0.32}$ & $6.6^{+7.2}_{-3.4}$ & $5.5^{+2.1}_{-1.4}$ & $0.41^{+0.62}_{-0.28}$ &  \\
VOS 24 & \textbf{G0} & $5920^{+120}_{-40}$ & $566.1^{+9.3}_{-9.0}$ & $2.418^{+0.075}_{-0.033}$ & $0.897^{+0.040}_{-0.023}$ & $2.67^{+0.16}_{-0.17}$ & $1.93^{+0.07}_{-0.14}$ & $4.7^{+1.3}_{-0.4}$ &  \\
VOS 26 & A2 & $\bf{9000^{+1000}_{-1000}}$ & $2402^{+92}_{-85}$ & $1.73^{+0.22}_{-0.32}$ & $2.03^{+0.19}_{-0.21}$ & $4.3^{+2.5}_{-1.6}$ & $3.12^{+0.85}_{-0.59}$ & $1.9^{+1.6}_{-1.0}$ &  \\
VOS 32 & \textbf{G0} & $5920^{+120}_{-40}$ & $1155^{+25}_{-24}$ & $3.005^{+0.070}_{-0.031}$ & $1.132^{+0.044}_{-0.027}$ & $3.50^{+0.23}_{-0.24}$ & $2.31^{+0.10}_{-0.16}$ & $2.96^{+0.80}_{-0.33}$ &  \\
\enddata
%\vspace{-0.8cm}
\tablecomments{VOS 1617 was observed with both IDS and EFOSC2 settings and was assigned a B1.5 and a B3 spectral type respectively. Given the better resolution of the IDS configuration and its smaller spectral type uncertainty we consider the B1.5 spectral type determination for this object in this work. Remarks column: `Cont' indicates the non-PMS contaminants discussed in Sect. \ref{S_contaminants}. `Dubious' marks those 13 sources described in Appendix \ref{S_dubious} with no detected H\textalpha{} emission and little IR excess. These stars have a less secure YSO nature than the other sources. `unclB[e]' indicates the four sources identified as of a FS CMa (unclassified B[e]) nature (see Appendix \ref{S_dubious}). `Young' shows the four sources that are younger than previously known stars of a similar mass (see Appendix \ref{S_dubious}). `T Tau' indicates that the source is inconsistent with M\textgreater{}$1.5$ M$_{\odot}$ and hence it likely belongs to the T Tauri regime.}
\end{deluxetable*}
\end{longrotatetable}

\begin{longrotatetable}
%\centerwidetable
\begin{deluxetable*}{lccccccl}
\tabletypesize{\footnotesize}
\tablewidth{0pt}
\tablecolumns{8}
\tablecaption{H\textalpha{} and H\textbeta{} observed equivalent widths (EW\textsubscript{obs}), line profiles, and the derived accretion rates and IR excesses. EW\textsubscript{corr} refers to the values which result from correcting EW\textsubscript{obs} from the typical line absorption EW of each spectral type (Sect. \ref{S_accre}). See Sect. \ref{S_accre} for a description on the derivation of the accretion rates. The amount of IR excess is shown in column L\textsubscript{IR}/L\textsubscript{*} (Sect. \ref{S_IR_excess}). \label{Table_accre}}
\tablehead{
\colhead{Name} & \colhead{H\textalpha{} EW\textsubscript{obs} [\AA{}]} & \colhead{H\textbeta{} EW\textsubscript{obs}} [\AA{}] & \colhead{EW\textsubscript{corr}} [\AA{}] & \colhead{Line profile} & \colhead{\.{M}acc} [M$_{\odot}$/yr] & \colhead{L\textsubscript{IR}/L\textsubscript{*}} & \colhead{Remarks}}
%\rotate
\startdata
VOS 4 & -3.30 &  & -13.00 & d & $(0.55^{+0.31}_{-0.20})\cdot10^{-7}$ & $0.316^{+0.067}_{-0.054}$ &  \\
VOS 7 &  &  &  &  &  & $0.39^{+0.13}_{-0.10}$ &  \\
VOS 15 & -8.70 &  & -17.75 & d & $(1.05^{+0.74}_{-0.46})\cdot10^{-7}$ & $0.58^{+0.17}_{-0.13}$ &  \\
VOS 20 & 3.23 &  & -4.98 & iP & $(2.02^{+0.71}_{-0.55})\cdot10^{-8}$ & $0.212^{+0.031}_{-0.011}$ &  \\
VOS 22 & -53.98 &  & -60.42 & d & $(0.11^{+0.25}_{-0.08})\cdot10^{-4}$ & $0.27^{+0.28}_{-0.12}$ &  \\
VOS 24 & 0.26 &  & -2.66 & d & $(1.30^{+0.46}_{-0.31})\cdot10^{-8}$ & $0.255^{+0.025}_{-0.031}$ &  \\
VOS 26 & -23.50 &  & -34.03 & d & $(0.16^{+0.22}_{-0.09})\cdot10^{-5}$ & $0.42^{+0.24}_{-0.14}$ &  \\
VOS 32 & 0.36 &  & -2.56 & rP & $(2.36^{+0.88}_{-0.60})\cdot10^{-8}$ & $0.586^{+0.030}_{-0.047}$ &  \\
\enddata
%\vspace{-0.8cm}
\tablecomments{H\textbeta{} EW\textsubscript{obs} values and H\textbeta{} line profiles are only presented when H\textbeta{} has emission and H\textalpha{} was not observed. Boldface type indicates that a value was derived from the H\textbeta{} line and not from the H\textalpha{} line. The line profiles are `s' for single-peaked emission, `d' for double-peaked, `rP' for regular P-Cygni, and `iP' for inverse P-Cygni. L\textsubscript{IR}/L\textsubscript{*} values marked with an asterisk likely have contamination from extended background emission (Sect. \ref{S_IR_excess}). Remarks column: `Cont' indicates the non-PMS contaminants discussed in Sect. \ref{S_contaminants}. `Dubious' marks those 13 sources described in Appendix \ref{S_dubious} with no detected H\textalpha{} emission and little IR excess. These stars have a less secure YSO nature than the other sources. `unclB[e]' indicates the four sources identified as of a FS CMa (unclassified B[e]) nature (see Appendix \ref{S_dubious}). `Young' shows the four sources that are younger than previously known stars of a similar mass (see Appendix \ref{S_dubious}). `T Tau' indicates that the source is inconsistent with M\textgreater{}$1.5$ M$_{\odot}$ and hence it likely belongs to the T Tauri regime.}
\end{deluxetable*}
\end{longrotatetable}

\bibliography{MyBib}{}
\bibliographystyle{aasjournal}

%% This command is needed to show the entire author+affiliation list when
%% the collaboration and author truncation commands are used.  It has to
%% go at the end of the manuscript.
%\allauthors

%% Include this line if you are using the \added, \replaced, \deleted
%% commands to see a summary list of all changes at the end of the article.
%\listofchanges

\end{document}